\AtBeginDocument{%
	\paperwidth=\dimexpr
	1in + \oddsidemargin
	+ \textwidth
	+ 1in + \oddsidemargin
	\relax
	\paperheight=\dimexpr
	1in + \topmargin
	+ \headheight + \headsep
	+ \textheight
	+ 1in + \topmargin
	\relax
	\usepackage[pass]{geometry}\relax
}

\RequirePackage{fix-cm}
\documentclass[smallextended, final, a4paper]{svjour3}       % onecolumn (second format)
\smartqed  % flush right qed marks, e.g. at end of proof
\usepackage[natbibapa]{apacite}
\usepackage{graphicx, array, blindtext}
\usepackage[colorlinks,citecolor=blue,urlcolor=blue]{hyperref}
\usepackage{amsmath} 
\usepackage{amssymb}
\usepackage{latexsym}
\usepackage{multirow}

\usepackage{verbatim}
\graphicspath{{images/}{./}} % To include images in other directories

\usepackage{url}
\usepackage[utf8]{inputenc}
\usepackage[space]{grffile}

\usepackage{tfrupee}
\usepackage{enumitem}\usepackage{graphicx, array, blindtext}

\usepackage{amsmath} 
\usepackage{amssymb}
\usepackage{latexsym}
\usepackage{multirow}

\usepackage{verbatim}
\graphicspath{{images/}{./}} % To include images in other directories

\usepackage{url}

\usepackage[utf8]{inputenc}

\usepackage[hang,flushmargin]{footmisc}

\usepackage[space]{grffile}

\usepackage{tfrupee}
\usepackage{enumitem}

\makeatletter
\renewcommand\section{\@startsection{section}{1}{\z@}%
	{-8\p@ \@plus -4\p@ \@minus -4\p@}%
	{6\p@ \@plus 4\p@ \@minus 4\p@}%
	{\normalfont\large\bfseries\boldmath
		\rightskip=\z@ \@plus 8em\pretolerance=10000 }}
\renewcommand\subsection{\@startsection{subsection}{2}{\z@}%
	{-8\p@ \@plus -4\p@ \@minus -4\p@}%
	{6\p@ \@plus 4\p@ \@minus 4\p@}%
	{\normalfont\normalsize\bfseries\boldmath
		\rightskip=\z@ \@plus 8em\pretolerance=10000 }}
\renewcommand\subsubsection{\@startsection{subsubsection}{3}{\z@}%
	{-4\p@ \@plus -4\p@ \@minus -4\p@}%
	{-1.5em \@plus -0.22em \@minus -0.1em}%
	{\normalfont\normalsize\bfseries\boldmath}}
\makeatother

\usepackage{etoolbox}

\makeatletter

\patchcmd{\NAT@citex}
{\@citea\NAT@hyper@{%
		\NAT@nmfmt{\NAT@nm}%
		\hyper@natlinkbreak{\NAT@aysep\NAT@spacechar}{\@citeb\@extra@b@citeb}%
		\NAT@date}}
{\@citea\NAT@nmfmt{\NAT@nm}%
	\NAT@aysep\NAT@spacechar\NAT@hyper@{\NAT@date}}{}{}

\patchcmd{\NAT@citex}
{\@citea\NAT@hyper@{%
		\NAT@nmfmt{\NAT@nm}%
		\hyper@natlinkbreak{\NAT@spacechar\NAT@@open\if*#1*\else#1\NAT@spacechar\fi}%
		{\@citeb\@extra@b@citeb}%
		\NAT@date}}
{\@citea\NAT@nmfmt{\NAT@nm}%
	\NAT@spacechar\NAT@@open\if*#1*\else#1\NAT@spacechar\fi\NAT@hyper@{\NAT@date}}
{}{}

\makeatother

\begin{document}

\title{A Patterns Based Approach for Design of Educational Technologies\thanks{\textit{Submitted to Educational Technology Research and Development Journal, Springer}\\\\This work was carried out as part of first author's doctoral thesis at IIIT Hyderabad, India and contains content from the thesis \citep{chimalakonda2017software}}}

%\thanks{Grants or other notes
%about the article that should go on the front page should be
%placed here. General acknowledgments should be placed at the end of the article.}

%\subtitle{Do you have a subtitle?\\ If so, write it here}

%\titlerunning{Short form of title}        % if too long for running head

\author{Sridhar Chimalakonda         \and
        Kesav V. Nori %etc.
}

%\vspace{-2cm}
%\authorrunning{Short form of author list} % if too long for running head

\institute{S. Chimalakonda \at
              Department of Computer Science \& Engineering\\
              Indian Institute of Technology, Tirupati\\
              %Tel.: +123-45-678910\\
              %Fax: +123-45-678910\\
              \email{ch@iittp.ac.in}           %  \\%             \emph{Present address:} of F. Author  %  if needed
           \and
           K.V. Nori \at
              Software Engineering Research Center\\
              International Institute of Information Technology Hyderabad, India\\
              \email{ch@iittp.ac.in}
}

%\date{Received: date / Accepted: date}
% The correct dates will be entered by the editor

\maketitle

\vspace{-1.5cm}

\begin{abstract}
	
Instructional design is a fundamental base for educational technologies as it lays the foundation to facilitate learning and teaching based on pedagogical underpinnings. However, most of the educational technologies today face two core challenges in this context: (i) lack of instructional design as a basis (ii) lack of support for a variety of instructional designs. In order to address these challenges, we propose a patterns based approach for design of educational technologies. This is in contrast with existing literature that focuses either on patterns in education or in software, and not both. The core idea of our approach is to leverage patterns for modeling instructional design knowledge and to connect it with patterns in software architecture. We discuss different categories of patterns in instructional design. We then present the notion of \textit{Pattern-Oriented Instructional Design} (POID) as a way to model instructional design as a connection of patterns (\textit{GoalPattern}, \textit{ProcessPattern}, \textit{ContentPattern}) and integrate it with \textit{Pattern-Oriented Software Architecture} (POSA) based on fundamental principles in software engineering. We demonstrate our approach through adult literacy case study (287 million learners, 22 Indian Languages and a variety of instructional designs). The results of our approach (both web and mobile versions) are available at \textit{http://rice.iiit.ac.in} and were adopted by \textit{National Literacy Mission Authority} of \textit{Government of India}.

\keywords{patterns; modeling; architecture; instructional design; software engineering; adult literacy}

\vspace{0.5cm}

\end{abstract}

%----------------------------------------------------------------------
% Pattern-Oriented Ontology Design - Journal Paper idea

\section{Introduction}\label{sec:introduction}

Education domain has been undergoing a major transformation in the last decade or so \citep{alper2009trends, hwang2011research, adams2017nmc, wu2012review}. On one hand, there is a significant surge on the use of educational technologies\footnote{We consider \textit{``educational technologies as a set of processes, techniques, methods and tools that facilitate learning and teaching based on well-established instructional designs."}} (such as game based learning \citep{tobias2014game}, MOOCs \citep{reich2015rebooting}, gesture based learning \citep{sheu2014taking}, augmented reality \citep{bower2014augmented} and so on) to facilitate learning and teaching. On the other hand, there is a significant fallout on the expectation of educational technologies ranging from terminological inconsistency \citep{bayne2015s} to the lack of effectiveness of instructional technology in classrooms \citep{venkatesh2014perceptions}. Several researchers have underlined the broken promises of educational technologies    \citep{bingimlas2009barriers,cuban2015dubious,spector2014handbook,spector2013emerging}. In addition, the community has identified several grand challenges in educational technologies \citep{woolf2013ai,fischer2014grand}, computing and information systems \citep{computing2003grand} in the context of education. The \textit{National Academy of Engineering} lists ``\textit{Advance personalized learning}" as a grand challenge of engineering\footnote{\url{http://www.engineeringchallenges.org/challenges/learning.aspx}}. In this paper, we are concerned about two core challenges that underlie most of these grand challenges:

\subsection{\textit{Challenge 1: Lack of Instructional Design as a Basis for Design of Educational Technologies}} 

Instructional Design\footnote{We consider \textit{instructional design as an underlying structure consisting of different aspects of instruction such as goals, process, content aimed at (i) providing a base for quality of instruction (ii) facilitating design of educational technologies}} has gained a significant role in the field of Technology Enhanced Learning as an underlying and complex discipline often involving multiple perspectives and connotations \citep{reigeluth2013instructional,reigeluth2013instructionalVolume2,reigeluth2009instructionalVolume3}. Merrill has defined instructional design as the practice of creating \textit{``instructional experiences which make the acquisition of knowledge and skill more efficient, effective, and appealing"} \citep{Merrill2012}. 

Berger defines instructional design as a \textit{``systematic development of instructional specifications using learning and instructional theory to ensure the quality of instruction”} \citep{berger1996definitions}. From the definition of Berger, instructional design acts as a basis for quality of instruction. In a similar line, Carroll emphasizes that quality of instructional design leads to quality of instruction \citep{carroll1963model}. Bednar et al. have stated that ``... effective instructional design is possible only if the developer has reflexive awareness of the theoretical basis underlying the design . . . [it] emerges from the deliberate application of some particular theory of learning" \citep{bednar1992theory}. This strong need to have a pedagogical basis for design of educational technologies has been emphasized in the literature by several researchers \citep{boyle2001towards,Govindasamy2001, goodyear2004patterns,garzotto2009critical,duffy2013constructivism}. A critical and detailed analysis of the need to bridge learning theories and technology enhanced learning environments is elaborated in \citep{lowyck2014bridging}. However, most of the educational technologies today lack instructional design basis leading to poor quality of instruction \citep{toyama2011there,lowyck2014bridging,spector2014handbook,reid2014categories,laurillard2013teaching,laurillard2013rethinking}. \newline

\textit{How to facilitate design of educational technologies with an instructional design basis?}

\subsection{\textit{Challenge 2: A Scale \& Variety of Instructional Designs and Educational Technologies}}

Instructional Design is used as an umbrella term that can refer to an entire discipline or as a process, art, science or technology \citep{brown2015essentials}. \textit{One size does not fit all} is a core principle that suits well for teaching and learning as every context is different because of aspects such as varied learning styles, varied instructional designs, varied learning environments and so on. There are over 100 instructional design models in the literature \citep{reigeluth2009instructionalVolume3} and several perspectives of instructional design. In one of the early works, a continuum of instructional strategies was presented with one end focusing on instructor-centered to the other end focusing on student-centered with a wide range of activities ranging from \textit{drill and practice} to \textit{projects} and \textit{inquiry} \citep{gustafson1997survey}. 

The ADDIE model involving \textit{analysis}, \textit{design}, \textit{development}, \textit{implementation} and \textit{evaluation} phases is one of the most widely used instructional design model that is applicable in several contexts \citep{kruse2002introduction}. Gagne's series of nine learning events \citep{gagne1974principles} has laid foundation for several instructional design models such as Dick and Carey model \citep{dick2001systematic} and Merrill's first principles of instruction \citep{Merrill2012}. Millwood summarizes over 25 learning theories in a concept map connecting the different facets of instructional design \citep{millwood2014design}. Gibbons \citep{gibbons2013architectural} presents eight views of instructional design (i) organizational view (ii) systems approach view (iii) design language view (iv) instructional systems design view (v) functional-modular view (vi) architectural view (vii) team process view (ix) operational principle view. Mizoguchi et al. have done an comprehensive survey of learning theories from instructional technology perspective and modeled them using ontologies \citep{mizoguchi2007inside}. The synthesis of literature on instructional design reveals that there is a strong need to support a variety of instructional designs during design of educational technologies. 

There has been extensive research on modeling instructional design for the last several years resulting in a plethora of educational modeling languages (EMLs) \citep{Martinez-Ortiz2007, Botturi2006,Botturi2008} such as poEML \citep{Caeiro2014}, PALO \citep{Rodriguez-Artacho2004}, Web COLLAGE \citep{villasclaras2013web} as a way to model and reuse aspects of instructional design. Sampson et al. presented an open access hierarchical framework for integrating open educational resources at different levels of granularity \citep{Sampson2014}. IMS-LD emerged as a standard for learning design \citep{ims2003ims} and then focus shifted to tools such as LAMS \citep{Dalziel2003} and LDSE \citep{Laurillard2013} that aim to support teachers. A vision paper aimed to create an approach that integrates most of these tools towards an integrated learning design environment \citep{Hernandez-Leo2013}. Despite this rapid progress, many researchers have pointed to several shortcomings of modeling and reusing instructional design such as complexity of authoring, lack of adequate tool support, interoperability and inability to support teachers \citep{Neumann2009}.

%\subsection{Challenge 3: Effort for Design of Educational Technologies}

%In this paper, we propose \textit{a patterns based approach for modeling instructional design and using that as the base for design of educational technologies}

\textit{How to facilitate design of educational technologies to support a variety of instructional designs?}

%\textit{The core contribution of this paper is a \textit{patterns based approach} for modeling a variety of instructional designs.}

%To address some of these concerns, patterns were proposed as a potential solution to capture best practices of teaching \citep{sharp2000pedagogical}.
%The core idea of \textit{patterns} and \textit{pattern languages} is the encapsulation, modeling and delivery of expert's knowledge and best practices to novices in a discipline. Essentially, patterns are derived from experiences and provide abstract representations of recurring solutions to recurring problems in a given context \citep{Alexander1977}. The roots of patterns are claimed to be in the field of architecture \citep{Alexander1977} and extensively practiced in software engineering mainly for improving quality of software design and facilitating reuse \citep{gamma1994design} \citep{buschmann2007pattern}. 

\textit{Section §\ref{sec:ch3GuidingPrinciples} of the paper discusses core guiding principles in computing towards a patterns based approach. We then present an overview of the source of patterns, their evolution and how to document them in Section §\ref{sec:ch3SourceEvolutionandStructureofPatterns}. How patterns can help scale and variety is discussed in Section §\ref{sec:ch3PatternsforScaleandVariety} followed by a patterns based approach for design of educational technologies in Section §\ref{sec:ch3APatternsBasedApproach}. Section §\ref{sec:ch3Pattern-OrientedInstructionalDesign} presents the idea of pattern-oriented instructional design with its sub sections proposing  patterns for goals in Section §\ref{sec:ch3GoalsPattern}, instructional process in Section §\ref{sec:ch3ProcessPattern} and instructional material in Section §\ref{sec:ch3ContentPattern}. We present Pattern-Oriented Software Architecture in Section §\ref{sec:ch3POSA} and discuss implementation of our approach in Section §\ref{sec:ch3Implementation}. We end the paper with conclusions and future work in Section §\ref{sec:ch3Conclusion}.}

\subsection{Patterns as a Solution} 

One interesting approach that has been undermined and largely unexploited in technology enhanced learning is the use of patterns and pattern languages for modeling instructional designs. The core idea of \textit{patterns} and \textit{pattern languages} is the encapsulation, modeling and delivery of expert's knowledge and best practices to novices in a discipline. Essentially, patterns are derived from experiences and provide abstract representations of recurring solutions to recurring problems in a given context \citep{Alexander1977}. The roots of patterns are claimed to be in the field of architecture \citep{Alexander1977} and extensively practiced in software engineering mainly for improving quality of software design and facilitating reuse \citep{gamma1994design, buschmann2007pattern}. 

There is also extensive work on patterns and pattern languages for different aspects of teaching and learning \citep{cristea2004designing,goodyear2004patterns}. The Pedagogy Patterns Project was a major effort to capture best practices in the area of teaching and learning as a way to document best advices for teachers and support quality of instruction \citep{sharp2000pedagogical} \citep{bergin2012pedagogical}. The E-LEN project is another initiative aimed at providing pedagogically-informed technology and experiences as pattern languages for new institutions mainly focusing on learning management systems \citep{avgeriou2003towards}. A pattern language for creative learning is presented in \citep{iba2010learning} and for adaptive learning in \citep{midgley2014goals}. Laurillard has created pedagogical patterns using a design science approach \citep{laurillard2012teaching}. Patterns and pattern repositories for person centered e-learning were proposed in \citep{derntl2004patterns}. Another direction was to mine patterns derived from practitioner workshops were documented in \citep{mor2014practical}. 

While existing literature on patterns emphasizes the need for capturing best practices in teaching and learning, the focus has been mostly on pedagogy and technology aspects are largely ignored. Even in research that considered technology, there is a huge gap between domain (teaching and learning) and technology patterns motivating further research. \textit{The key focus in this paper is to leverage the potential of patterns for modeling a scale and variety of instructional designs and use them as a base for design of educational technologies.
}

\section{Guiding Principles} 
\label{sec:ch3GuidingPrinciples}

To support design of educational technologies for \textit{scale} and \textit{variety}, the proposed approach is based on the following underlying principles.

It is a dire necessity to improve flexibility and re-usability while reducing complexity during the design of educational technologies and some of these concerns are tackled by the software engineering community through a set of fundamental principles \citep{ghezzi2002fundamentals}. One principle that is of interest to this paper is the notion of \textit{\textbf{separation of concerns}} that helps in handling different dimensions of a system while improving re-usability and reducing complexity \citep{dijkstra1982role,greenfield2004software}. This principle can be used to separate concerns in the form of layers either horizontally or vertically; views \citep{kruchten19954+}; modules \citep{parnas1972criteria}; aspects \citep{kiczales1997aspect}; patterns \citep{greenfield2004software}; features \citep{kang1990feature} and so on. \textit{\textbf{Modularity}} is a specialization of this principle that deals with separating software into components \citep{parnas1972criteria} while \textit{\textbf{Abstraction}} is another specialization that hides complexity of the system enabling designers to focus on specific concerns \citep{ghezzi2002fundamentals}. Domain-driven design is another fundamental principle that tries to address complexity by emphasizing domain as the basis for software design \citep{evans2004domain}. We apply these principles throughout the paper for systematically modeling instructional design in the form of patterns. We discuss these principles further in Section  §\ref{sec:ch3Pattern-OrientedInstructionalDesign}. In the next section, we will summarize some key aspects related to patterns and get into the crux of our pattern-oriented design approach.

%------------------------------
\section{Source, Evolution and Structure of Patterns} 
\label{sec:ch3SourceEvolutionandStructureofPatterns}

\begin{figure*}[t]
	\centering
	\includegraphics[width=1\linewidth,clip=true, trim = 0mm 10mm 0mm 10mm]{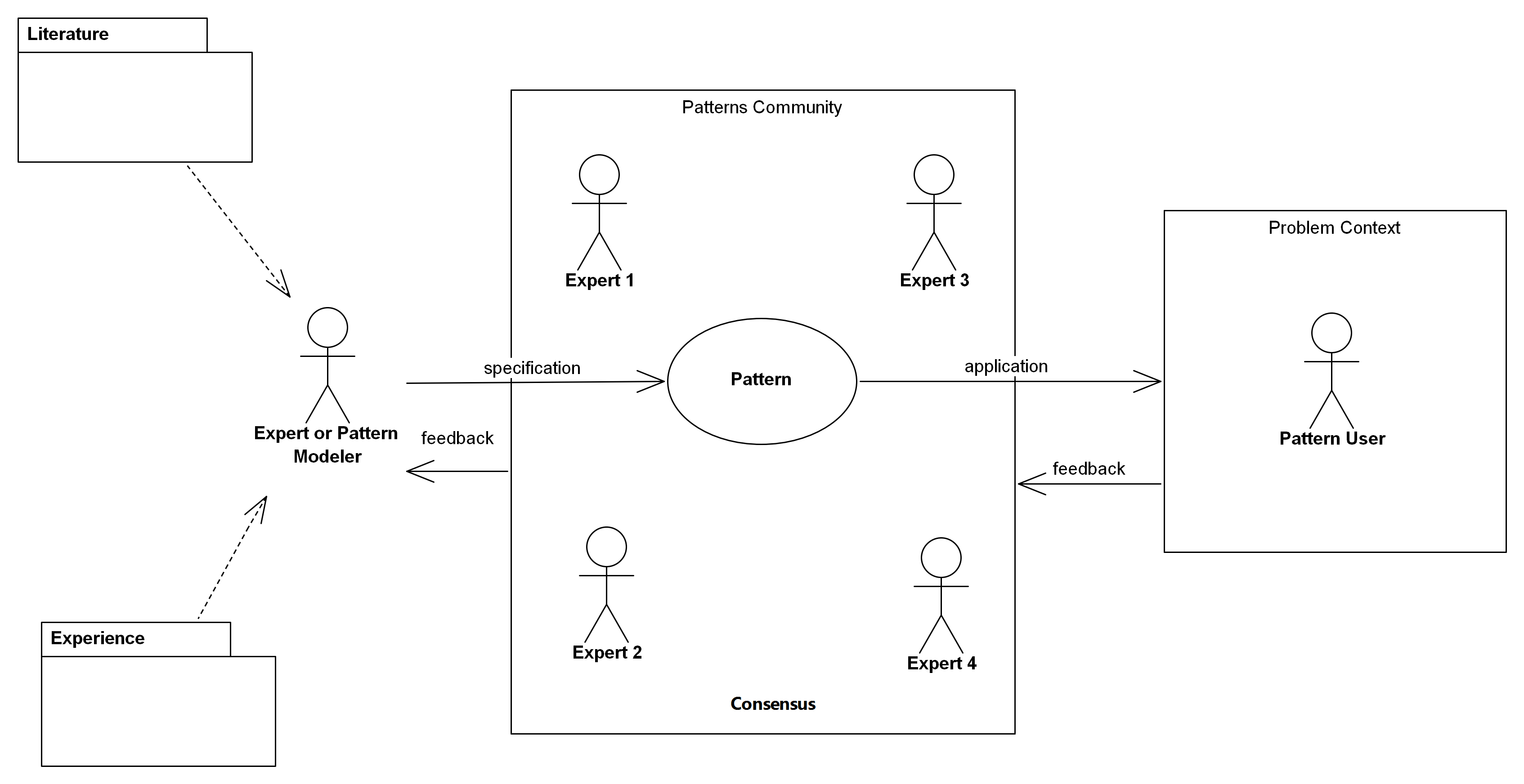}
	\caption{Overview of patterns life cycle}
	\label{fig:ch3PatternsLifeCycle}
	%\vspace{1cm}
\end{figure*}

\textbf{Source and Evolution of Patterns} - During the professional journey of an expert in a field, the expert encounters several recurring problems and different ways of solving those problems. When a group of experts in a particular field communicate, discuss, debate and document their experience, they realize that certain solutions work and do not work in particular contexts. This leads to the idea that the primary source of patterns is literature or experience, either own or documented experience in the form of best practices and guidelines. Granularity of a pattern is another critical aspect that has to be considered during the design of patterns. For example, in the field of software engineering, there are coding idioms (a kind of patterns), design patterns, architectural patterns, each representing an increasing granularity of abstraction. Pattern languages help in connecting these patterns by describing the relationships between them and how they can be integrated to address specific problems. The agreement on whether a particular knowledge is pattern or not usually comes through a consensus among a group of experts in the particular field. The Hillside group has been sponsoring a series of conferences along with workshops named Pattern Languages of Programs (PLoP) since 1994 \citep{berczuk1994finding}. These venues provide a forum for pattern authors to gather, discuss, learn and document patterns. This series has led to development of communities of patterns such as \textit{AsianPLoP}, \textit{EuroPLoP}, \textit{ScrumPLoP} in order to create patterns with a consensus from local communities. A pattern typically goes through the phases of \textit{discovery}, \textit{specification}, \textit{validation}, \textit{application} and \textit{maintenance} and is continuously revised based on updated pattern knowledge \citep{derntl2005patterns}. Figure §\ref{fig:ch3PatternsLifeCycle} shows a typical process during pattern development. In the domain of education also, patterns have mainly emerged from participatory pattern workshops \citep{mor2012participatory}.

\begin{figure*}[t]
	\centering
	\includegraphics[width=1\linewidth,clip=true, clip=true, trim = 0mm 30mm 0mm 20mm]{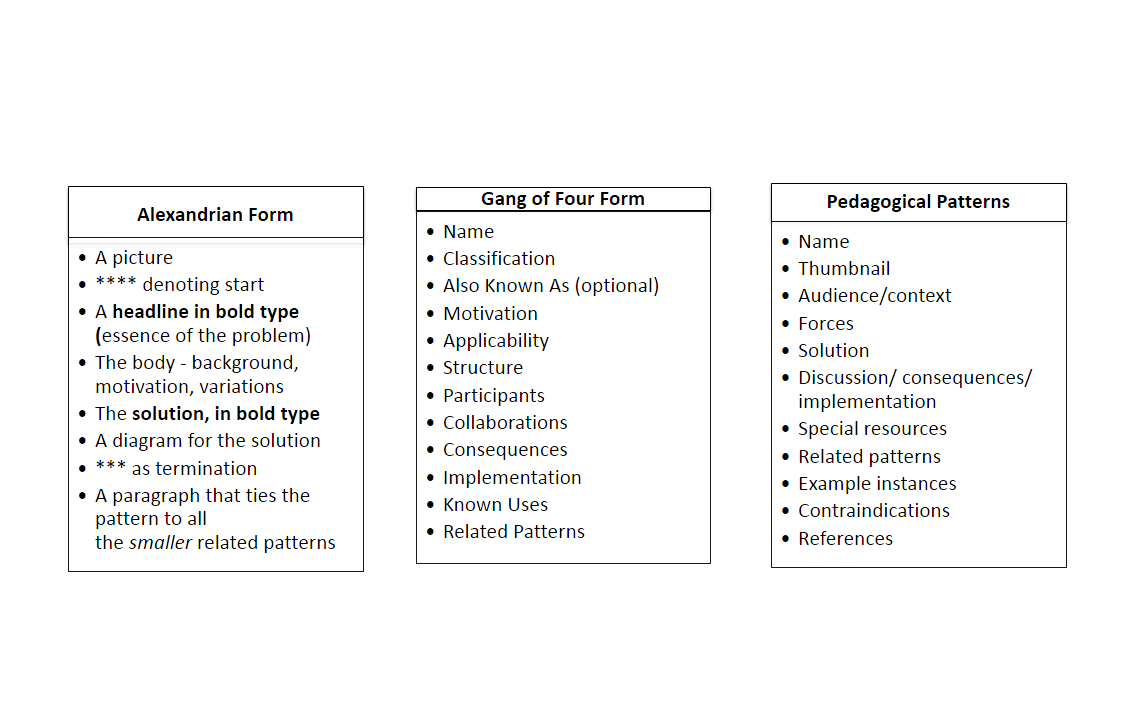}
	\caption{Diversity of structure of patterns}
	\label{fig:ch3PatternForms}
\end{figure*}

In essence, patterns generally emerge from literature or experience, and are generally specified by an expert or pattern modeler. Once the pattern is specified, it is exposed to the community for review from experts in the field for a consensus of the pattern. This pattern is applied by pattern users in varied problem contexts and they provide feedback leading to update of pattern. 

\begin{figure*}[t]
	\centering
	\includegraphics[width=1\linewidth,clip=true, clip=true]{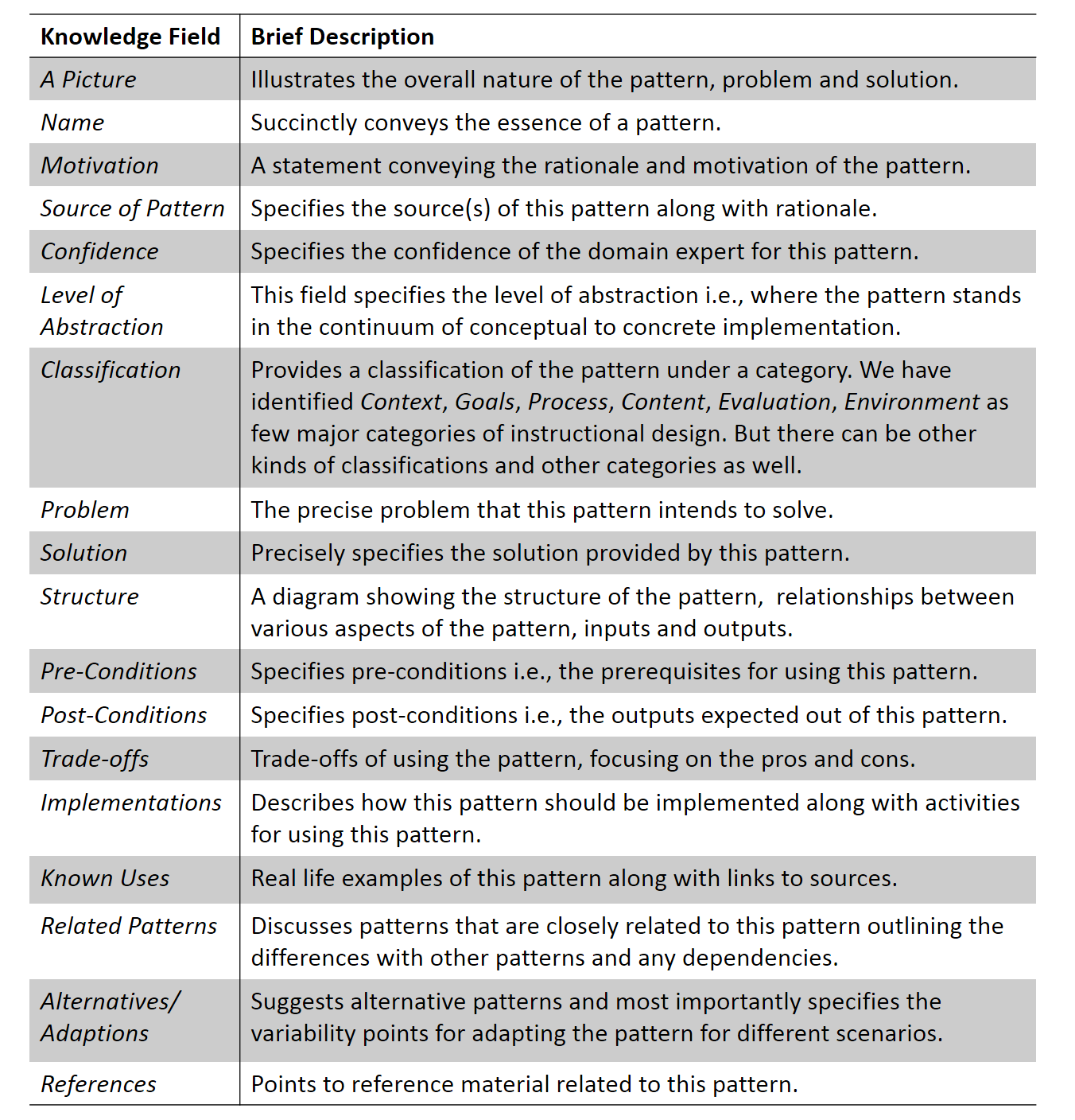}
	\caption{A detailed pattern structure}
	\label{fig:ch3PatternStructure}
\end{figure*}

\begin{figure*}[t]
	\centering
	\includegraphics[width=1\linewidth]{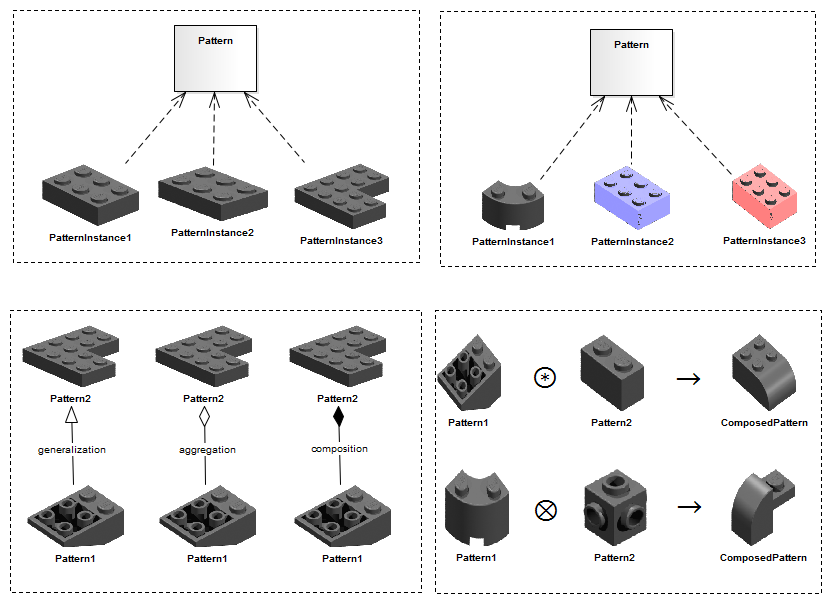}
	\caption{Some pattern variations, relationships and compositions}
	\label{fig:ch3PatternVariations}
\end{figure*}

\textbf{Structure of Patterns} - There are patterns everywhere and at different levels of granularity expressed in different ways by different experts in different communities \citep{meszaros1998pattern}. This presents the challenge of what constitutes pattern knowledge and how to capture it. Traditionally, patterns are captured as a description of \textit{context}, \textit{problem}, \textit{solution}. However, several researchers have proposed different structures for capturing pattern knowledge. Figure §\ref{fig:ch3PatternForms} shows examples of commonly used pattern structures from the domains of software engineering \citep{gamma1994design} and pedagogy \citep{2014apedagogicalproject}. Alexandrian form is the most commonly used format for representing patterns \citep{Alexander1977}. Using this form, Alexander has documented a pattern language comprising of 253 patterns for the domain of towns, buildings and construction \citep{Alexander1977}. The 23 Gang of Four (GoF) patterns are the most commonly used patterns in the domain of software design \citep{gamma1994design} and use the meta data as shown in Figure §\ref{fig:ch3PatternForms}. The Pedagogical Patterns project uses the structure shown in Figure §\ref{fig:ch3PatternForms}. These are just a few examples of pattern structures used by different communities. A pattern language for representing patterns itself has emerged from the community \citep{meszaros1998pattern}. 

%\begin{figure*}[b]
%	\centering
%	\includegraphics[width=1\linewidth,clip=true, clip=true, trim = 0mm 60mm 0mm 50mm]{"./figures/ch3/ch3PatternStructure/Slide1"}
%	\caption{A detailed pattern structure}
%	\label{fig:ch3PatternStructure}
%\end{figure*}

\section{Patterns for Scale and Variety} 
\label{sec:ch3PatternsforScaleandVariety}

% A million different implementations page 96 posa 5 and a billion differnt implementations posa 5 page 330?

Figure §\ref{fig:ch3PatternStructure} shows a detailed pattern structure derived from existing pattern representations. The key idea of this pattern structure is to capture as much information as possible such that this metadata could be used for searching and managing patterns and pattern repositories. In addition to the textual information as shown in Figure §\ref{fig:ch3PatternStructure}, patterns could be represented visually focusing on \textit{structure} as well as \textit{behaviour}. However, one critical requirement for patterns in this paper is to facilitate domain experts to model pattern knowledge without overburden.  

In his seminal work on patterns, Alexander emphasizes that a solution in pattern can be used \textit{``a million times over without ever doing it the same time twice"}. We extend this notion to include not just the solution but also the problem, its variations, representations of the pattern, different aspects of the solution and its variants to facilitate scale and variety. 

There are several possibilities of creating variants of patterns. The simplest case being that a pattern can be instantiated multiple times as shown on the left hand side of Figure §\ref{fig:ch3PatternVariations} or instantiated with simple variations as shown on the right hand side. In Figure §\ref{fig:ch3PatternVariations}, we show a simple example of different kinds of possible relationships between two patterns in a domain. If we change these relationships, we get a different view of the pattern in context leading to a variation. Similarly, patterns can be composed using different operators as shown in Figure §\ref{fig:ch3PatternVariations}. Changing the operators results in new composed patterns and variations. If we move beyond just two patterns and consider a large number of patterns, then different ways of composition with multiple operators leads to different ways of modeling the domain. Consider the case of an instructional design with 10 patterns, then different compositions of these patterns can lead to several instructional design variants. This leads to creation of varied instructional designs catering to the needs of specific requirements. The essence of this discussion is to emphasize that modeling domain in terms of \textit{patterns} facilitates systematic creation of several variants, which is one of the primary goals of this paper. We will elaborate more on these patterns with examples throughout the paper. But formalizing the representation and composition of patterns is beyond the scope of this paper and is outlined as future work. 

%\begin{figure*}
%	\centering
%	\includegraphics[width=1\linewidth]{"./figures/ch3/ch3PatternInstancesBasicVariations"}
%	\caption{Simple pattern variations}
%	\label{fig:chBackgroundIPCL}
%\end{figure*}

%\begin{figure*}
%	\centering
%	\includegraphics[width=1\linewidth]{"./figures/ch3/ch3PatternRelationships"}
%	\caption{Some pattern relationships}
%	\label{fig:chBackgroundIPCL}
%\end{figure*}

\section{A Patterns Based Approach} 
\label{sec:ch3APatternsBasedApproach}

The notion of patterns has its roots in the field of Architecture \citep{Alexander1977} but was adopted in other disciplines such as software engineering \citep{gamma1994design} and interaction design \citep{borchers2001pattern} among others. Whilst there exists several definitions and views, \textit{patterns are primarily concerned with the idea of finding recurring solutions to recurring problems in a certain context}. According to Alexander, the emphasis has to be on pattern languages that facilitate the assembly of patterns in order to create numerous possible solutions, rather than patterns themselves \citep{Alexander1977}. Buschmann et al. have distilled existing literature on patterns and proposed the following uses \citep{buschmann1996pattern,buschmann2007pattern}:

%in their Pattern Oriented Software Architecture (POSA) series

\begin{itemize}
	
	\item \textit{Capturing, Documenting and Communicating Experience}
	\begin{itemize}
		
		\item Patterns document existing best practices built on tried and tested design experience
		\item Patterns identify and specify abstractions that are above the level of single objects, classes, and components
		\item Patterns provide a common vocabulary and shared understanding for design concepts
		\item Patterns are a means of documenting software architectures
		\item Patterns capture experience in a form that can be independent of specific project details and constraints, implementation paradigm, and often even programming language
		
	\end{itemize}
	
	\item \textit{Construction of Systems}
	
	\begin{itemize}
		\item Patterns support the construction of software with well-defined properties
		\item Patterns help in building complex and heterogeneous software architectures. Every pattern provides a predefined set of components, roles and relationships between them
	\end{itemize}
	
\end{itemize}

\begin{figure*}[t]
	\centering
	\includegraphics[width=1\linewidth]{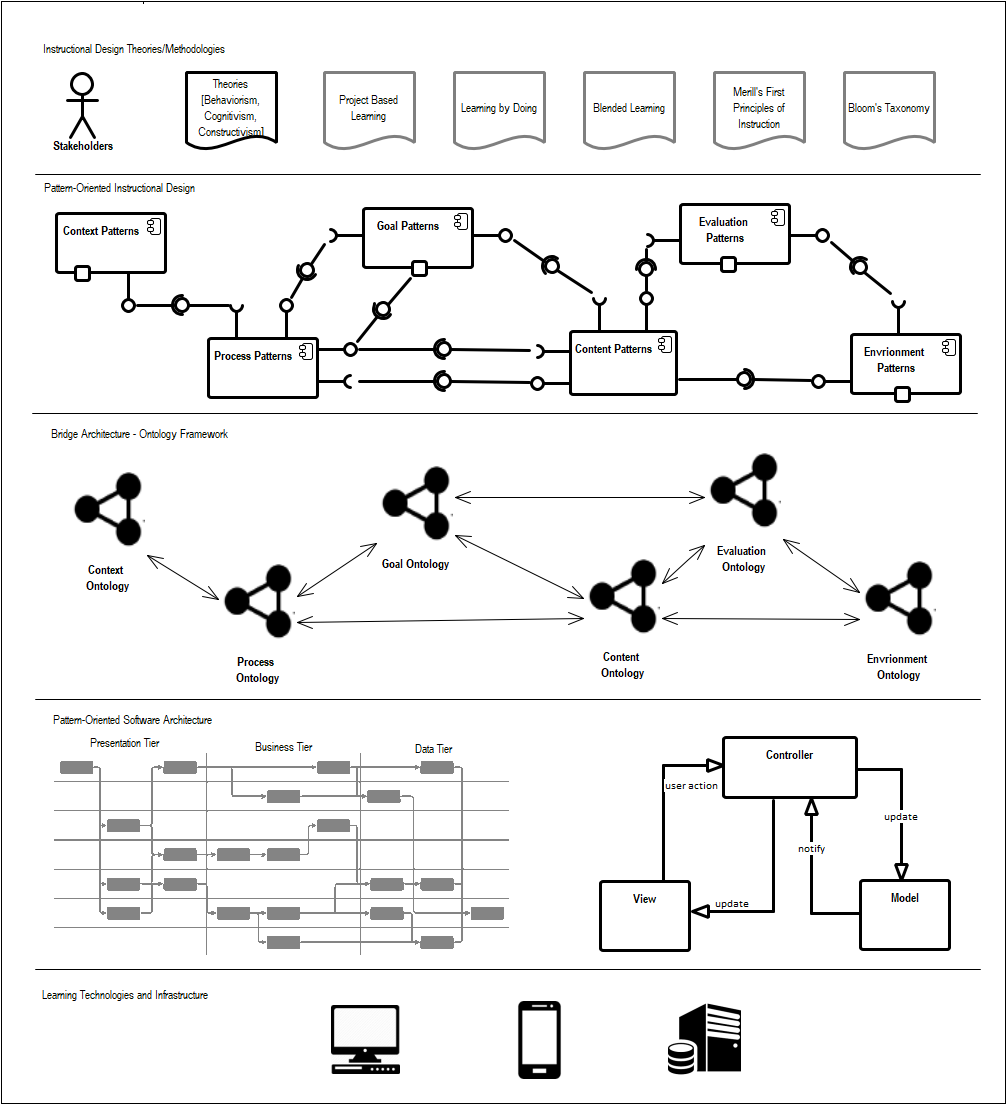}
	\caption{Architecture of patterns based approach}
	\label{fig:ch3ArchtiectureofPatternsBasedApproach}
\end{figure*}

In this paper, we apply the idea of patterns primarily for (i) \textit{capturing experience} (ii) \textit{providing instructional design as basis for educational technologies} (iii) \textit{facilitating reuse during design of educational technologies}. Specifically, we use patterns for modeling domain (instructional design) and software. We also look at patterns as a central way to encapsulate commonly understood knowledge of experts (instructional designers, software architects) and facilitate use of this experience by naïve professionals. This is extremely important in a discipline like Technology Enhanced Learning (TEL) with huge scarcity of expert teachers at all levels of education. Figure §\ref{fig:ch3ArchtiectureofPatternsBasedApproach} presents the core architecture of our proposed patterns based approach to design of educational technologies. This architecture stems from fundamental principles in software engineering and integrates multiple architecture styles (Layered, Domain-Driven and Component-Architecture). This architecture shows a holistic perspective (top-down and bottom-up) and tries to integrate patterns in both domain as well as software through five layers. 

We briefly explain the core design principles and different layers of our approach in the next sections.  \newline

\noindent \textbf{Core Design Principles}

\textbf{Separation of Concerns (SoC) Principle} -
SoC principle was applied in \citep{caeiro2006separation,Rodriguez-Artacho2004} for modeling instructional design but an analysis of this work reveals that this principle was applied either in modeling instructional design or in creating technology and not both, which we aim to address through our approach. As shown in Figure §\ref{fig:ch3ArchtiectureofPatternsBasedApproach}, we continuously applied SoC principle at different levels (low-level, high-level and pattern approaches) in instructional design and software as a way to address complexity, evolution and reusability of instructional design. We also applied the notion of abstraction as an underlying principle of SoC to separate technology specific aspects from technology independent aspects through bridge architecture as shown in Figure §\ref{fig:ch3ArchtiectureofPatternsBasedApproach}.

\textbf{Design for Reuse and Customization} -
It is extremely effort intensive to design educational technologies for scale and variety. Learning from the negative impact of ad hoc reuse in software engineering, our approach in Figure §\ref{fig:ch3ArchtiectureofPatternsBasedApproach} is explicitly designed for systematic reuse emphasizing a patterns-based approach. Every aspect of this architecture is modeled with explicit interfaces using required dependencies, provided services, open ports, assemblers and connectors. This ensures that each pattern makes its assumptions and capabilities explicit enabling systematic reuse.

\textbf{Domain Driven Architecture Design} - 
According to this architecture style, the design of software systems is essentially a realization of the underlying domain that is modeled by experts in that domain. In Figure §\ref{fig:ch3ArchtiectureofPatternsBasedApproach}, POSA is driven by POID emerging from domain. By design, POID and POSA are closely mapped via patterns in instructional design and software engineering. In addition, this architecture follows a layered style with several layers from the top dealing with domain and layers from the bottom dealing with technology. Even though the layers shown in the diagram seem to be fixed, the number of layers can be increased or decreased based on the application type, quality attributes of desired system, and technology constraints among others. Bridge architecture (message bus architecture style) defined via well defined interfaces allows better communication between the layers.

%Elaborate here as much as you can from software engineering perspective by giving specific references and more... May be not necessary now till you know Prof's stand

\begin{itemize}
	\item [A] \textit{Instructional Design/Methodologies} - lays a pedagogical foundation for design of educational technologies - 
	Educational experts have long emphasized that developing educational technologies without strong instructional basis is futile and can lead to poor quality of instruction \citep{Govindasamy2001,laurillard2013rethinking}. Most of the educational technologies today require huge effort from teachers in configuring the technologies rather than on focusing instruction \citep{laurillard2013constructionist}. This is further aggravated with a huge dearth of qualified teachers. The first layer from the top in Figure §\ref{fig:ch3ArchtiectureofPatternsBasedApproach} focuses on providing a pedagogical foundation for design of educational technologies. We rely on well-established principles and practices from a pedagogical perspective in this paper and specifically focus on how we can structure these practices towards systematic design of educational technologies. 
	For the case of adult literacy in this paper, we rely on Improved Pace and Contents of Learning (IPCL) approach \citep{2003IPCL} as a pedagogical foundation. This pedagogy is further integrated with other commonly accepted approaches such as Merrill's principles of instruction from teaching or instructional process perspective \citep{Merrill2012} and Bloom's taxonomy from learning perspective. The rationale for these principles is presented in Section §\ref{sec:ch3GoalsPattern} and Section §\ref{sec:ch3ProcessPattern} of patterns. 
	
	\item [B] \textit{Pattern-Oriented Instructional Design} - The goal of this layer to model instructional design using patterns (Section §\ref{sec:ch3Pattern-OrientedInstructionalDesign})
	\item [C] \textit{An Ontology Based Modeling Framework} - acts a bridge between domain and software platforms - The key purpose of this bridge layer is to connect from instructional design (domain) to educational technologies (software) through a common interface. We used ontologies as the primary mechanism to represent knowledge from patterns and to connect with the rest of the software architecture \citep{Chimalakonda2013d}. We extended the existing literature on instructional design, and proposed an ontology based framework called ID\textit{ont} for systematically modeling instructional design \citep{Chimalakonda2013d}.
	
	\item [D] \textit{Pattern-Oriented Software Architecture} - models the software architecture of educational technologies using patterns - In their seminal work that inspired successful use of patterns in software engineering, Buschmann et al. emphasize that the primary purpose of architectural patterns is to create a fundamental structural organization schema for software systems \citep{buschmann1996pattern}. In our approach, instructional architecture patterns derived from POID and software architecture patterns that drive POSA provide this structure for instructional design and educational technologies. We use common interfaces to bridge these different types of patterns at different levels of abstraction \textit{(instructional architecture patterns $\rightarrow$ instructional design patterns $\rightarrow$ software architecture patterns $\rightarrow$ design patterns)}. Essentially, POSA represents the architecture of educational technologies. In the fourth layer and on the left hand side of Figure §\ref{fig:ch3ArchtiectureofPatternsBasedApproach}, we have a three-tier architecture of a TEL system that implements two important architectural patterns from \citep{buschmann1996pattern} (i)  From Mud to Structure (Layers) – allows controlled decomposition of overall system (ii) MVC and several design patterns (\textit{Strategy}, \textit{Composite}, \textit{Factory} and so on). On the right hand side is an MVC architectural pattern that is derived from instructional architecture patterns and instructional design patterns in POID.
	
	\item [E] \textit{Technologies, Platform and Infrastructure} provides the delivery platform with a set of technologies and tools. In this layer, a set of technologies and tools are designed to support the proposed architecture and facilitate the semi-automatic development of \textit{e}Learning Systems.
	
\end{itemize}

We will elaborate POID and POSA in the rest of the paper. 

\begin{comment}
A high level overview of our approach is shown in Figure 1. The core idea of a patterns based approach is to design solutions relying on well known and commonly accepted practices. On the left hand side, the focus is on modeling instructional design based on patterns and on the right hand side, software patterns are used during the design of educational technologies. A detailed overview of our approach is shown in Figure 2. 
\end{comment}

%-------------------------------
\section{Pattern-Oriented Instructional Design} 
\label{sec:ch3Pattern-OrientedInstructionalDesign}

The changing landscape of educational technologies requires a diversified range of instructional designs catering to multiple perspectives and views from a diversified range of stakeholders \citep{reigeluth2013instructional}. This is primarily because there is no \textit{``one-size-fits-all"} solution for all needs. In fact, even the problem itself varies depending on who is viewing it, where does it come from, how critical is it? what are the short term and long term needs? how to address them? their capabilities and resources available at that point of time. This presents a definite need for systematic instructional design such that it can be flexibly modified as per changing requirements. Several researchers have tried to address these concerns through a number of Educational modeling languages (EMLs) \citep{Caeiro2014,Rodriguez-Artacho2004,villasclaras2013web,Botturi2008,koper2005introduction,Dalziel2003,Hernandez-Leo2013,hernandez2014ldshake}. Despite significant research, modeling instructional design remained an open research problem with several challenges \citep{Hernandez-Leo2013,Neumann2009,goddard2015has,burgos2015critical}. Our analysis of the state-of-the-art in learning design also revealed that the goal of end-to-end automation (from pedagogy to technology) through tools has resulted in too generic, too complex specifications leading to slow progress in this field and is in line with existing research \citep{laurillard2013constructionist,laurillard2013rethinking}.

Instead, we learn from the community and attempt to present a pointed approach towards design of educational technologies for a family of instructional designs in the context of adult literacy in India. \textit{Our work fundamentally deviates from the state-of-the-art as we focus on not one instructional design but on a family of similar but distinct instructional designs}.

%One interesting approach that has been undermined and largely unexploited in TEL is the use of patterns and pattern languages for modeling instructional design. There has been a strong recognition for use of patterns in the community but mainly from a learning sciences perspective \citep{borchers2001pattern}\citep{bergin2012pedagogical}. The Pedagogical Patterns Project provides a huge collection of patterns focusing on pedagogy and does not explicitly consider the use of technology and software patterns \citep{2014apedagogicalproject}. Patterns for person centered e-learning were proposed in \citep{derntl2005patterns} based on action research but are not designed for scale and variety. Pattern languages for different aspects of educational design \citep{Goodyear2010} also ignored the notion of TEL. Mor has discussed the notion of identifying design narratives as a basis for capturing design knowledge, developing design patterns based on that and showing the practical use of these patterns through design scenarios \citep{mor2013snap}. Laurillard has proposed teaching as a design science with the idea of supporting computational and collaborative building of pedagogies \citep{laurillard2012teaching}. In an attempt towards supporting teachers in practical usage of patterns, a collection of patterns have been presented in \citep{mor2014practical}. 

In our analysis of literature, despite immense work, we have observed that patterns and pattern languages are still not widely used either in instructional design or in TEL because of several reasons including:
\begin{itemize}
	\item Existing approaches focus on patterns and pattern languages either for communication or engineering purposes and not both
	\item Current approaches focus on patterns mainly from a pedagogical perspective rather than their structure and provide minimal support towards design of educational technologies
	\item Lack of bridge between domain patterns (instructional design) and technology patterns (software)
\end{itemize}

To summarize, existing approaches focus on patterns either in domain or in software and not both. Most importantly, none of the existing works focus on \textit{scale} and \textit{variety}, which is the main goal of this paper. Paquette has summarized the extensive work and progress in the field of instructional design and concluded the strong need for researching and applying ontological engineering and software methods for instructional design and engineering \citep{paquette2014technology}. On the other hand, from a software engineering perspective, domain engineering is a critical activity to address complexity, reuse and evolution needs of software systems \citep{taylor2009software}. In this paper, we take a cue from Apel et al. \citep{apel2013feature} and Czarnecki et al. \citep{czarnecki2000generative} and consider domain as an area of knowledge:

\begin{itemize}
	\item that covers the desired requirements of the systems in that area
	\item includes a set of concepts and terminology understood by practitioners in that area
	\item and includes the knowledge of how to build software systems (or parts of software systems) in that area
\end{itemize}

Modeling and structuring domain is a fundamental step that acts as a basis towards facilitating reuse and in this paper we are concerned with the domain of instructional design. We propose Pattern-Oriented Instructional Design as \textit{a domain engineering activity} towards design of educational technologies based on instructional design. POID aims at designing a solution in the problem domain in the language of instructional designers and teachers. The key input for POID comes from pedagogies or learning methodologies that provide a basis for TEL. The purpose of this layer is two-fold (i) to structure instructional design for reuse (ii) to facilitate flexible modeling of instructional designs such that educational technologies based on these instructional designs have a pedagogical basis and are prepared for facilitating automation. 

\begin{figure*}[t]
	\centering
	\includegraphics[width=1\linewidth, trim = 0mm 20mm 0mm 0mm]{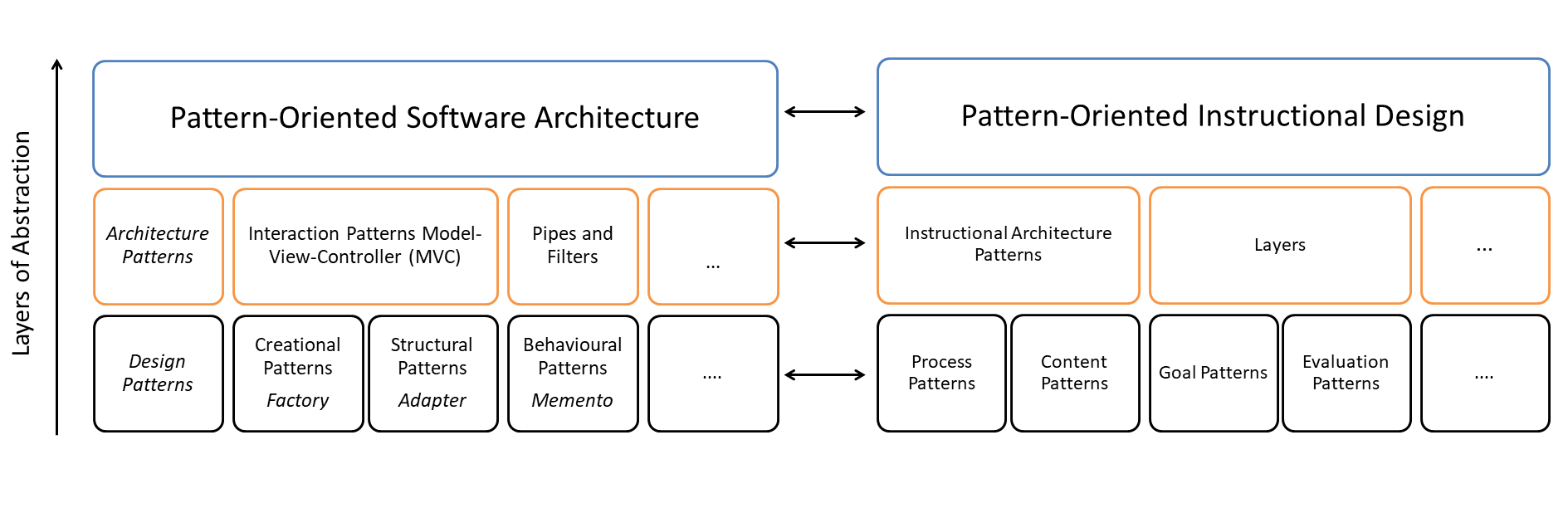}
	\caption{Comparision of Pattern-Oriented Software Architecture and Pattern-Oriented Instructional Design}
	\label{fig:ch3ComparisionofPOIDandPOSA}
\end{figure*}

We base the design of POID on Pattern-Oriented Software Architecture (POSA) \citep{buschmann2007pattern} to model instructional design aspects as patterns. Figure §\ref{fig:ch3ComparisionofPOIDandPOSA} shows a high level diagram of how POID corresponds to POSA with different levels of granularity. For example, on the left hand side, we have design patterns at the bottom and then architecture patterns. On similar lines, we have instructional design patterns at the bottom and instructional architecture patterns on top of them. 

\begin{figure*}
	\centering
	\includegraphics[width=1\linewidth,height=7.5cm]{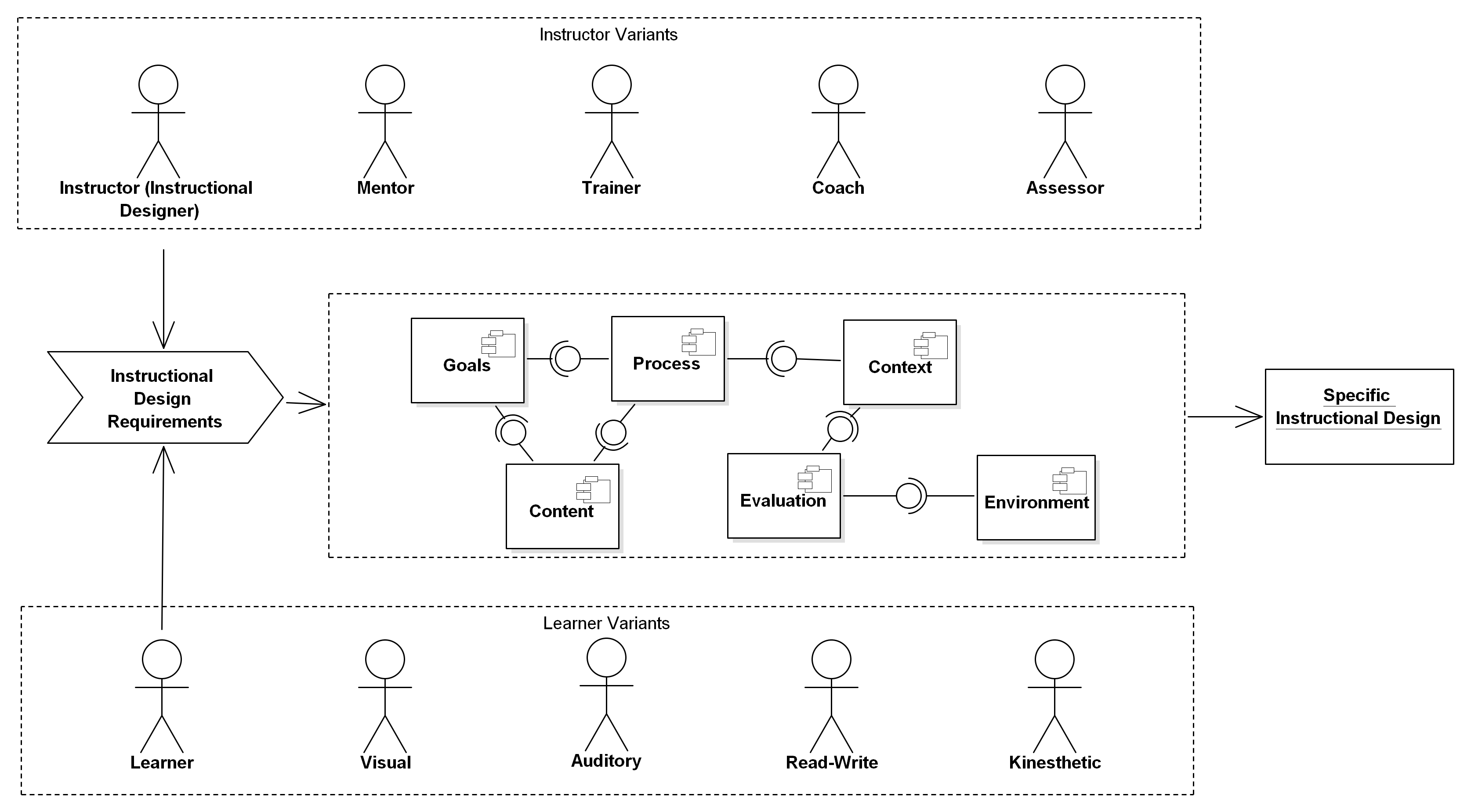}
	\caption{An overview of Pattern-Oriented Instructional Design}
	\label{fig:ch3POIDOverview}
\end{figure*}

Figure §\ref{fig:ch3POIDOverview} shows an overview of Pattern-Oriented Instructional Design. The core instructional design requirements stem from different kinds of instructors and learners, which are then used as input to compose different patterns such as \textit{goals}, \textit{process}, \textit{content} and so on to create a specific instructional design.
\textit{A Pattern-Oriented Instructional Design is an integration of patterns at instructional architecture level and instructional design patterns towards designing a specific instructional design for a specific set of educational requirements}. Here, we consider instructional architectural patterns as high-level organizing structures that address recurring high-level problems in instructional design. For example, Can we have a pattern that allows different evaluations for the same instructional goals? These patterns are essentially an integration of instructional design patterns, which focus on a specific aspect such as goals or process and so on. One popular example of an architectural pattern for interactive systems in software engineering is the Model-View-Controller (MVC) that allows flexible design of user interfaces \citep{krasner1988description}. Can we have these kinds of patterns for instructional design? It is here our POID approach is a direction integrating patterns at a higher level of abstraction.

\begin{figure*}[t]
	\centering
	\includegraphics[width=1\linewidth, trim = 0mm 30mm 0mm 30mm ]{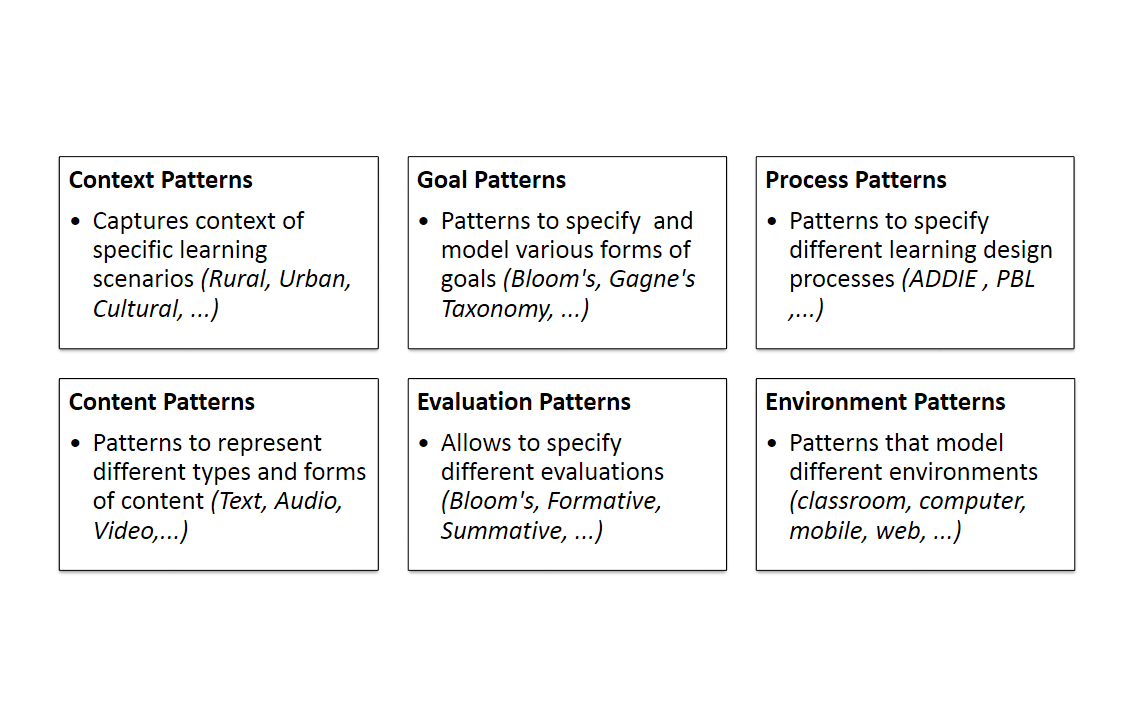}
	\caption{A classification of patterns in instructional design}
	\label{fig:ch3ClassificationofPatterns}
\end{figure*}

The POID process starts by understanding instructional design and then progresses towards a detailed analysis of the instructional design to identify patterns in the domain. Then these patterns are continuously refined and relationships between them are identified and established. Figure §\ref{fig:ch3ClassificationofPatterns} shows a classification of patterns in POID into categories of \textit{Context}, \textit{Goals}, \textit{Process}, \textit{Content}, \textit{Evaluation} and \textit{Environment}. This list is primitive and is designed such that it can be adapted and extended by instructional designers to support evolution. Each of these categories have patterns that aim to address a particular aspect of instructional design. For example, \textit{Goals} in instructional design can be represented in various forms such as Bloom`s or Gagne`s taxonomy or ABCD technique, and each of them can have their associated patterns, from which the teacher or instructional designer chooses patterns for their particular context. In our approach, every pattern provides and requires a set of interfaces that clearly define the boundaries of that pattern and how that pattern communicates with the rest of the patterns. The focus in this paper is not just on patterns or pattern categories, but on integration of these patterns towards specific instructional designs. 

Figure §\ref{fig:ch3ArchtiectureofPatternsBasedApproach} shows two possible instructional architecture patterns; one that integrates goals and evaluations (goal provides an interface $\leftrightarrow$ evaluation requires an interface). The core idea of this pattern is to provide a flexible architecture that allows changing of goals and evaluations in an instructional design with less effort. Another possible pattern could be an integration of processes and content. This kind of abstraction leads to POID, which in itself is a technology independent solution in the language of problem domain. However, it is important to note that patterns were never proposed to be specific and complete solutions rather provide a basic structure of a generic solution to a family of problems \citep{buschmann2007pattern}, which have to be further adapted and implemented for a specific context. Finally, Alexander hints that the problem of scale and variety can be addressed using patterns in his seminal work \citep{Alexander1977} as:

\textit{``The elements of this language are entities called patterns. Each pattern describes a problem that occurs over and over again in our environment, and then describes the core of the solution to that problem, in such a way that you can use this solution a million times over, without ever doing it the same way twice."}

This emphasizes that patterns provide an opportunity to find a generic solution to a problem and then find several customized solutions to a family of similar problems. Even though patterns are generic in nature, they have to be discovered from literature and experience for a particular context. In this paper, we use our decade-long experience in the case of adult literacy in India. \textit{We conducted a workshop in collaboration with Tata Consultancy Services and in association with NLMA to discover our patterns with the community towards a consensus}\footnote{at TCS, Hyderabad, India in 2011}. The workshop consisted of directors of State Resource Centers representing officers who are responsible for creating instructional process and material for teaching adult illiterates based on local requirements. Specifically, we introduced our patterns for instructional process and content and incorporated their feedback. We also held interviews with the directors of SRCs to figure out their use of technology for adult literacy especially to gather what kinds of technologies have been accepted and what are the hindrances to the use of technologies. One key learning that came out of that workshop is a strong requirement that the technology has to be customized as per local needs. One director specifically noted that ``the instructional process that you follow in a state like \textit{Bihar} and the one in a state like \textit{Tamilnadu} differ and the technology has to be adapted as per varied needs". In addition to our experience, we also rely on commonly accepted approaches such as Bloom`s taxonomy \citep{Anderson2001}, Merrill`s first principles of instruction \citep{Merrill2012} as a basis for the patterns proposed in this paper. 

In the following section, we detail some of the instructional design patterns we discovered in the context of adult literacy in India. 

\subsection{A pattern for modeling goals} 
\label{sec:ch3GoalsPattern}

We assume that any instruction is \textit{goal-driven}; making it critical to explicitly state instructional goals. Several terminologies such as ``learning goals", ``learning outcomes", ``learning objectives" have been in use to discuss goals. But the crux is to explicitly state and represent these goals such that they become clear to different stakeholders such as teachers, learners, policy makers and so on. In our case, for design of educational technologies while adhering to instructional design principles, they can also help in tracking the progress of learners, evaluation and in helping teachers to improve instructional strategies. We advocate a \textit{goal-driven approach} throughout this paper as it is critical and essential for any instructional design.  

\begin{figure*}[!b]
	\centering
	\includegraphics[width=0.8\linewidth]{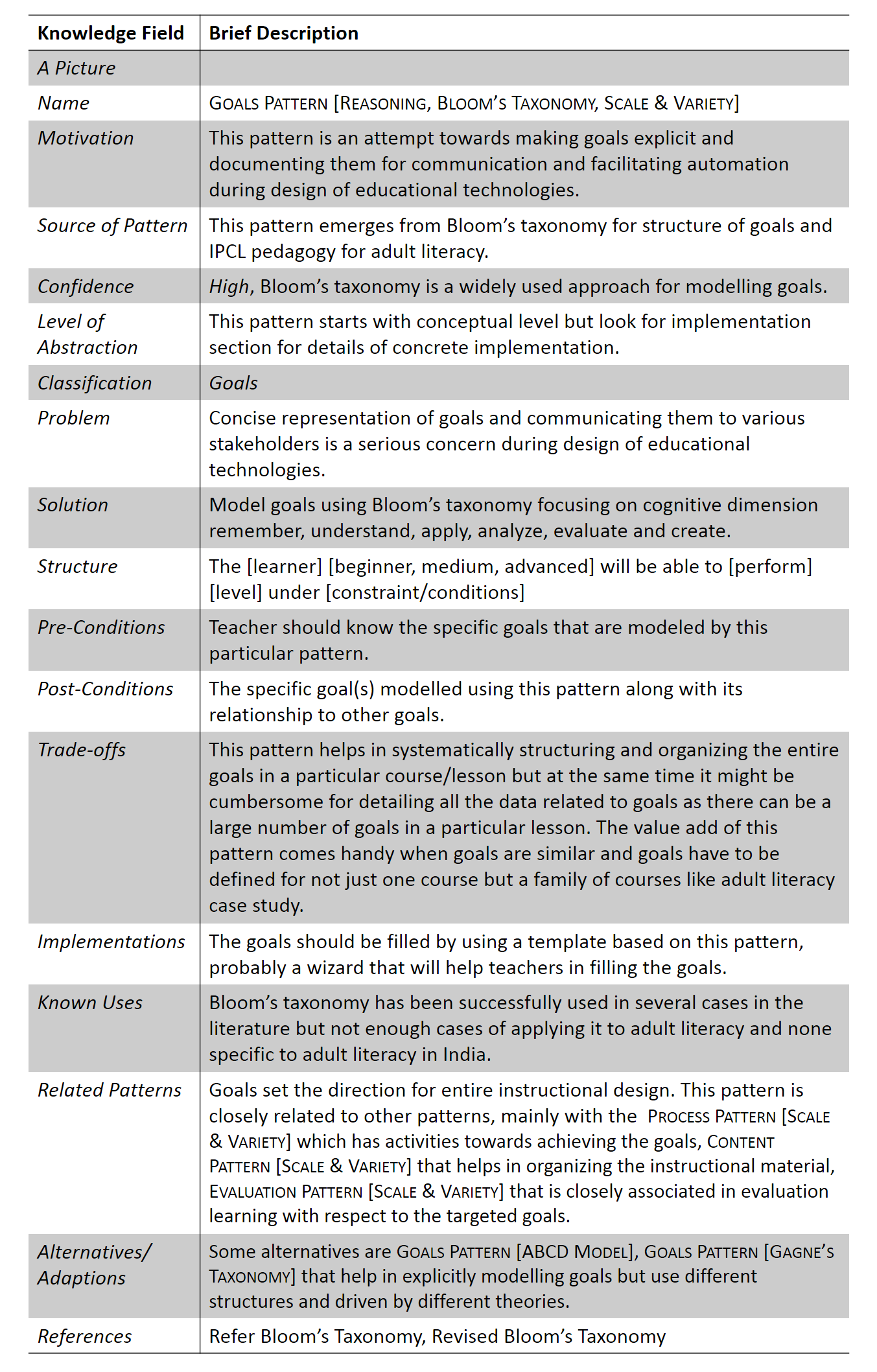}
	\caption{A sample structure and partial instance of \textit{GoalsPattern}}
	\label{fig:ch3GoalsPattern}
\end{figure*}

For example, if we consider the representation of goals in instructional design, several variations are possible. In the simplest case, a goal can be represented using Bloom`s taxonomy and several instances of this goal can be created. If we extend the scenario and think of two goals as part of instructional design, the two goals can be associated in several ways. 

\begin{figure*}[!b]
	\centering
	\includegraphics[width=1\linewidth]{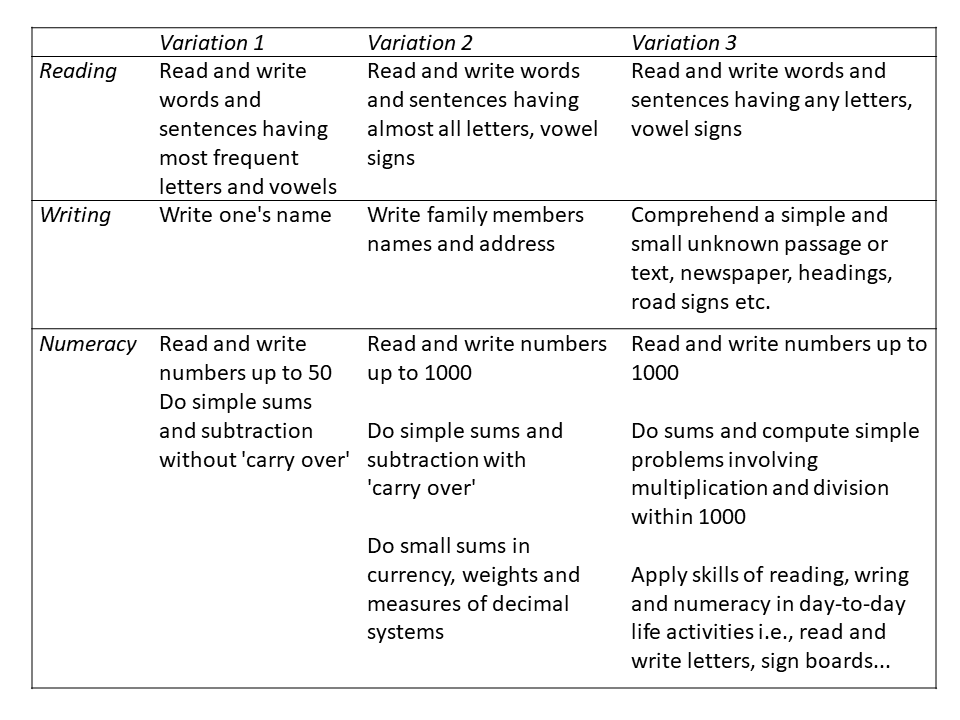}
	\caption{Examples of goals in adult literacy}
	\label{fig:ch3GoalsPattern1}
\end{figure*}

The crux of most commonly used Bloom`s taxonomy is to provide a classification for organizing educational objectives based on thinking models and to facilitate better communication among stakeholders in education \citep{Anderson2001}. This taxonomy has evolved since its inception but was originally divided into three domains: \textit{cognitive}, \textit{affective}, \textit{pyschometer} with the first two domains focusing on acquiring knowledge and attitude and the third domain on skills to put that knowledge to constructive use \citep{Anderson2001}. Each of these domains are further divided into levels that indicate progress \textit{``from simple to complex and concrete to abstract"} \citep{Anderson2001}. In revised Bloom`s taxonomy, goals are organized in the increasing order of complexity in six levels: \textit{remember}, \textit{understand}, \textit{apply}, \textit{analyze}, \textit{evaluate}, and \textit{create} from a cognitive domain perspective. ABCD model proposed by R. Mager is another commonly used framework for writing learning objectives \citep{mager1962preparing}. In this model, a learning objective consists of four components: [\textit{audience}- who] will be able to [\textit{behavior}-perform] [\textit{condition}- constraints] [\textit{degree} - level of quality]. The point of this discussion is to emphasize that there are several ways of modeling goals motivating the need for several patterns for goals in different contexts. In the context of this paper, based on these inputs and our experience of designing \textit{e}Learning Systems for adult literacy in India, we propose a pattern to model goals based on these inputs. Figure §\ref{fig:ch3GoalsPattern} shows our proposed pattern for modeling goals based on Bloom`s taxonomy. Figure §\ref{fig:ch3GoalsPattern1} shows some examples of goals for adult literacy based on IPCL methodology and Bloom's taxonomy.  

%An example of a goal from IPCL methodology and its modeling using Bloom`s taxonomy is as follows:

%\textit{``Reading aloud, with normal accent, and at a speed of 30 words a minute, a simple passage on a topic of interest to the\ learner"}

These goals can be refined for varied lessons as part of instructional design and can be customized for specific multiple Indian languages.

%\begin{itemize}
%	\item The learner will be able to \textit{recognize} three syllables {\hin "म", "क", "न"} in an existing paragraph
%	\item The learner will be able to \textit{recognize} three syllables {\tel "క", "థ", "ూ"} in an existing paragraph
%\item The learner will be able to \textit{recognize} three syllables {\ "த்", "ப்", "ழ்"} in an existing paragraph	
% \item The learner will be able to \textit{read} the syllables aloud at a speed of 30 words per minute
% \textbf{}\item The learner will be able to \textit{remember} three syllables {\hin "म", "क", "न"} 
%\end{itemize}

\begin{comment}
\begin{itemize}
\item Reading aloud, with normal accent, and at a speed of 30 words a minute, a simple passage on a topic of interest to the learner.
\item Silent reading at 35 words a minute, of small paragraphs in simple language.
\item Reading with understanding the road signs, posters, simple instructions and newspapers/broadsheets designed for neo-literates.
\item Ability to follow simple written passages relating to one's working and living environment.
\item Copying, with understanding, 7 words a minute.
\item Taking dictation at 5 words a minute.
\item Writing with proper spacing and alignment.
\item Writing independently, short letters and applications, and filling in forms of day-to-day use to the learner.
\item Reading and writing numerals 1-100.
\end{itemize}
\end{comment}

%What is this pattern? Explain

%What are two variations of this pattern?

%Structure and behaviour of the pattern...

\subsection{A pattern for modeling instructional processes} 
\label{sec:ch3ProcessPattern}

Instructional process is one of the critical aspects of instructional design as it facilitates the fulfillment of goals through a systematic process. However, most of the times it is not explicitly modeled by making it difficult for design of educational technologies. In this section, we will look at a commonly accepted way of teaching in the context of adult literacy in India based on IPCL methodology \citep{2003IPCL} and present a structure for organizing that knowledge into a pattern. We discuss the instructional process in detail along with teaching philosophy as it forms the basis for a pattern that could be instantiated thousands of times for all Indian languages and dialects.

\begin{figure*}[!b]
	\centering
	\includegraphics[width=1\linewidth, trim= 0mm 10mm 0mm 10mm]{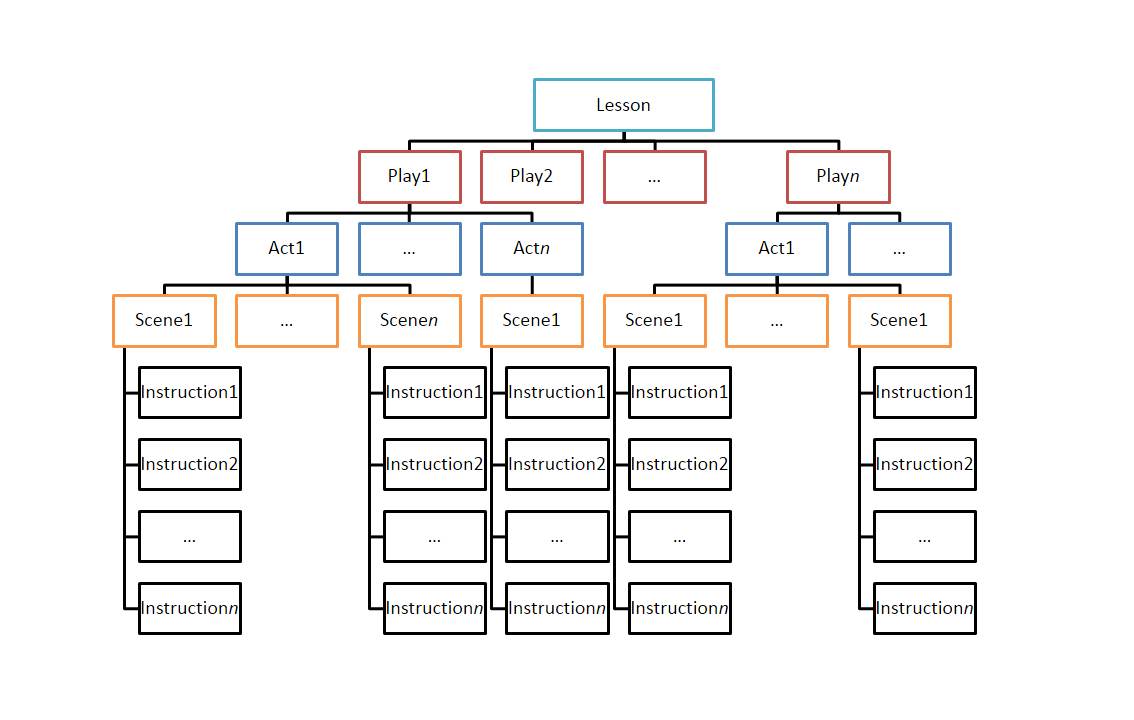}
	\caption{Structure of instructional process}
	\label{fig:ch3StructureofInstructionalProcess}
\end{figure*}

%\begin{figure*}
%	\centering
%	\includegraphics[width=1\linewidth, trim= 0mm 10mm 0mm 10mm]{"./figures/ch3/ch3/Slide4"}
%	\caption{Part of an example \textit{play} in \textit{Hindi} language based on \textit{ProcessPattern}(\textit{pasi})}
%	\label{fig:ch3ProcessPatternExample}
%\end{figure*}
%\begin{figure*}
%	\centering
%	\includegraphics[width=1\linewidth, trim= 0mm 10mm 0mm 10mm]{"./figures/ch3/ch3/Slide7"}
%	\caption{Part of an example \textit{play} in \textit{Telugu} language based on \textit{ProcessPattern} (\textit{pasi})}
%	\label{fig:ch3ProcessPatternExampleTelugu}
%\end{figure*} 

\begin{figure*}
	\centering
	\includegraphics[width=1\linewidth]{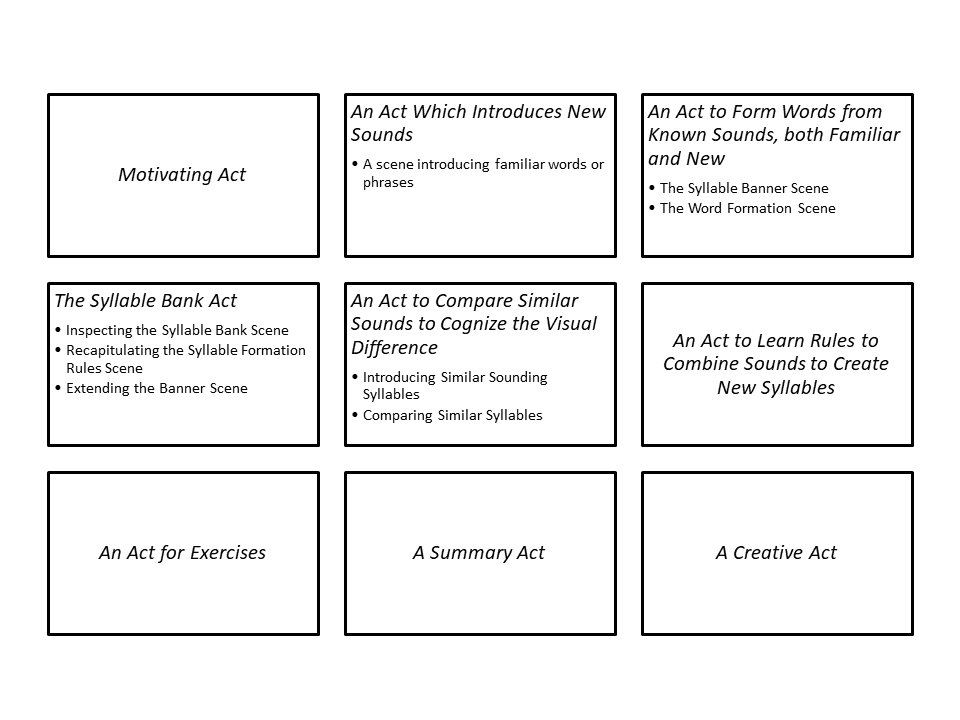}
	\caption{Example \textit{acts} in adult literacy instructional design}
	\label{fig:ch3ProcessPatterns}
\end{figure*}

Figure §\ref{fig:ch3StructureofInstructionalProcess} shows organizational structure of a \textit{lesson} using the \textit{pasi} pattern \citep{chimalakonda2017software}. The number and order of the plays, acts, scenes, instructions is not strictly fixed even though guidelines can be framed. For example, the first \textit{play}, \textit{act} and \textit{scene} focus on providing motivation to the learner and the last \textit{instruction} might be a summary of what has been learnt so far in a particular lesson. Figure §\ref{fig:ch3ProcessPatterns} shows few examples of \textit{acts} and some \textit{scenes}. In this example, there are several \textit{acts} each having its respective goals, and consisting of specific \textit{scenes} and further \textit{instructions}. For example \textit{Act4} deals with the goal of teaching how to form new words from syllables with two \textit{scenes} illustrating how new words are formed from already learnt syllables. This \textit{ProcessPattern} has several sources of variation for instruction process. The variations can be the \textit{number} of plays, acts, scenes, instructions; the \textit{order} of them, the specific play, act, scene or instruction, the content used and other aspects of instruction providing customization for scale and variety.

 %Figure §\ref{fig:ch3ProcessPatternExampleTelugu} shows the same example in Figure §\ref{fig:ch3ProcessPatternExample} with variations introduced for \textit{Telugu} language.   

While the \textit{ProcessPattern} provides a goal-driven structure for modeling a pedagogy at a high-level, it does not include a strong philosophical attitude of instructional process and does not specify how these goals have to be achieved. It is here we analyzed the literature in instructional design and found that there are several perspectives of instructional design. There are also several ways of modeling instructional processes based on different instructional design theories or methodologies. Merrill has analyzed existing instructional design models and proposed that the following fundamental principles are critical to any instructional design \citep{Merrill2012}. 

\begin{itemize}
	\item Activation principle - reaching out to what students know
	\item Application principle - exercising their new knowledge 
	\item Integration principle - accumulating or integrating what they have learnt recently with what was learnt in the past 
	\item Demonstration principle - showing how this new knowledge can be used
	\item Task Orientation principle - getting students to solve problems
\end{itemize}

Each of these principles (activities) are repeatedly used in a specific order in the instructional process to fulfill goals. In addition to these principles, Merrill also proposed a deeper sub-cycle \textit{structure–guidance–coaching–reflection} that strengthens these activities. For example, a structure has to be provided for the learner as part of instruction while applying activation principle and necessary guidance has to be given to the learners during a demonstration activity. 

Generally, the \textit{ProcessPattern} involves some or all of these principles at different levels of granularity but the application of these principles becomes explicit for tasks at \textit{instruction} level. So, for example \textit{Scene1} of \textit{Act2} introduces words that are familiar to the learners essentially involving activation principle whereas learners have to use application principle in \textit{Scene2} of \textit{Act5} to form new words from existing syllables. Similarly, other instructions in the instructional process can be mapped to principles. 

\subsection{A pattern for modeling instructional material} 
\label{sec:ch3ContentPattern}

\begin{figure*}[!t]
	\centering
	\includegraphics[width=1\linewidth, trim = 0mm 30mm 0mm 20mm]{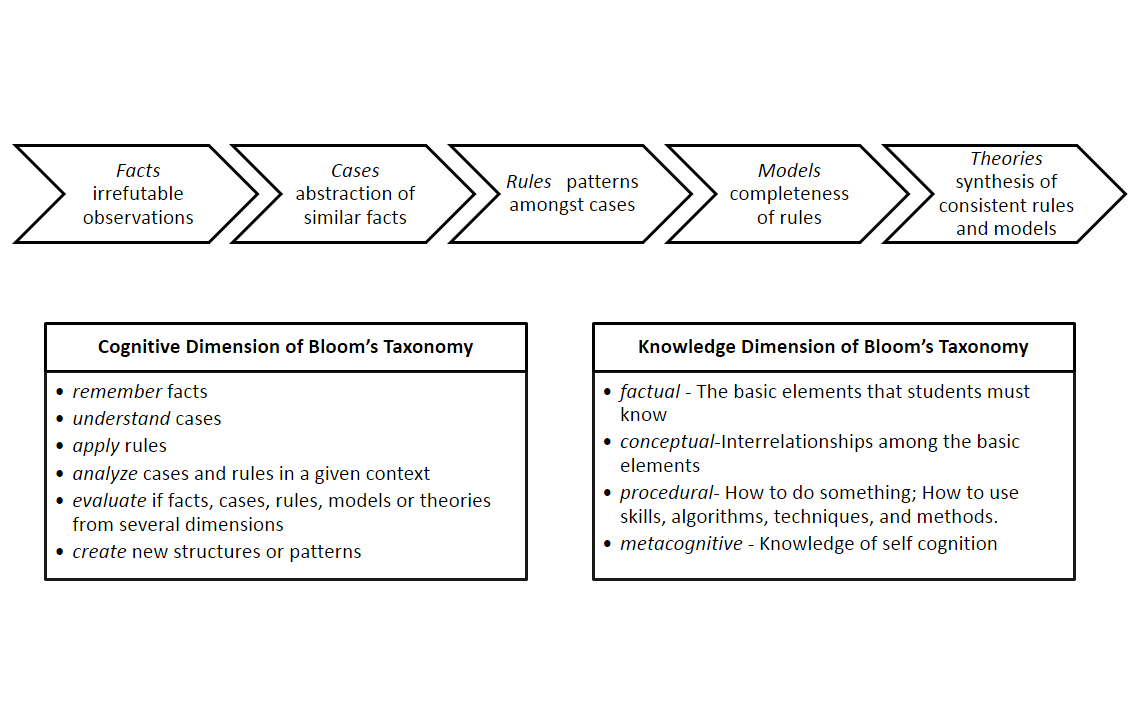}
	\caption{Pattern structure and mapping to cognitive and knowledge dimensions of Bloom's taxonomy}
	\label{fig:ch3MappingFCRMTtoBloom}
\end{figure*}

Instructional material is at the center of instruction and we discovered a pattern for content as shown in Figure §\ref{fig:ch3MappingFCRMTtoBloom}. This pattern is primarily derived from IPCL, scientific method but most importantly driven by our future need to facilitate reasoning in the subject. Can learners provide rationale and reasoning for their answers? 

\begin{figure*}[t]
	\centering
	\includegraphics[width=1\linewidth, trim = 0mm 10mm 0mm 10mm]{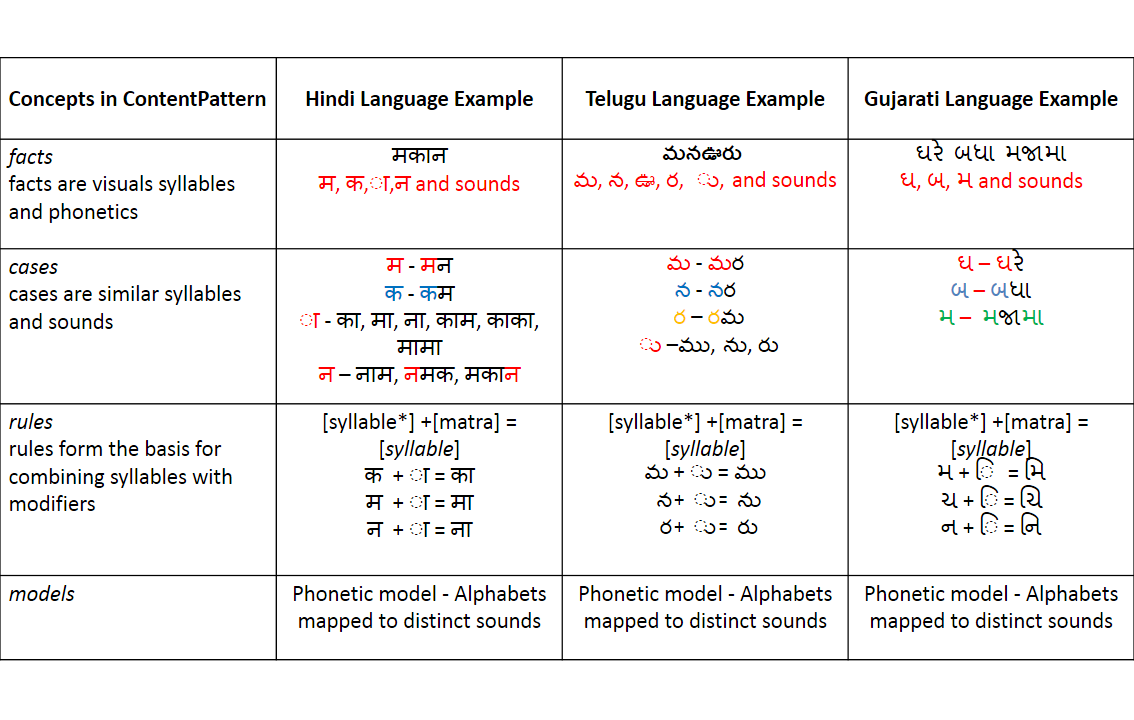}
	\caption{Example of \textit{ContentPattern} instances for \textit{Hindi}, \textit{Telugu} and \textit{Gujarati} languages}
	\label{fig:ch3ContentPatternHindiTeluguGujaratiExamples}
\end{figure*}

Figure §\ref{fig:ch3MappingFCRMTtoBloom} shows the progression of content from simple facts to be remembered to foundations in the subject. The mapping of this pattern to cognitive and knowledge dimensions of Bloom's taxonomy is also shown in the figure. We discussed the foundations of this pattern during its formative stages in \citep{Chimalakonda2012e}. Figure §\ref{fig:ch3ContentPatternHindiTeluguGujaratiExamples} shows instances of the \textit{ContentPattern} in \textit{Hindi}, \textit{Telugu} and \textit{Gujarati} languages.    

The \textit{ProcessPattern} and \textit{ContentPattern} are closely connected and can be considered as an instructional architecture pattern. In this pattern, teaching is structured, not simply as a sequence of lectures, but as a sequence of \textit{models} as needed by a \textit{theory} to be taught. Here, we consider teaching as a \textit{play} uncovering each \textit{model} as the instruction unfolds, with learner participation! Each \textit{model} in turn would be a sequence of topics. Uncovering a \textit{model} is done in a succession of \textit{acts}, with each \textit{act} uncovering a topic pertinent to some \textit{model}. Each \textit{act} is presented as a succession of \textit{scenes}, with each \textit{scene} focusing on \textit{cases} or \textit{rules}. Finally, each \textit{scene} is delivered as a succession of basic \textit{instructions}, which uncover and play with \textit{facts}. 

\section{Pattern-Oriented Software Architecture} 
\label{sec:ch3POSA}
%Teaching as a step-wise refinement 

Software architecture deals with most of the design decisions concerned with \textit{structure}, \textit{behavior}, \textit{interaction}, \textit{qualities}, and \textit{implementation} of a software system \citep{taylor2009software}. In this paper, we are interested in identifying the variants that are possible in each of these aspects to support the desired requirements of scale and variety in educational technologies. In general, architectural patterns and reference architectures provide a baseline and a set of guidelines for creating specific software architectures tapping the variabilities of multiple related systems in a particular domain \citep{taylor2009software}. At a high-level, customization of software can be done at design level, requiring explicit modifications to architecture and design of software and also from a user perspective, requiring configurations \citep{morch1997three}. Morche \citep{morch1997three} lists three common ways of tailoring software applications from literature: 

The users can \textit{customize} the application by configuring a set of pre-defined options mostly related to user interface and existing functionality. These options can range from themes or colors in the user interface to turn off or turn on features of the existing system. For example, an instructor might configure the theme of the \textit{e}Learning System based on local context and culture, evaluation in terms of multiple choice or fill in the blanks and so on.

In \textit{integration}, users can have configuration options to integrate functionality that is outside the system. For example, a teacher might want to integrate learning management system like \textit{Moodle} into the current system. This might require tweaking of components, extending interfaces and fixing interoperability issues. A GLUE-architecture was proposed to support integration of tools into virtual environments \citep{alario2013glue}. 

When additional new features have to be added, the system has to be \textit{extended} by adding new code, re-writing and sometimes even re-designing some parts of the system. Extending the system becomes an expensive activity if the current system is not designed for extensibility. Extensions can emerge from new requirements from domain or because of new techniques and technologies available to design software. For example, if a new accreditation rule requires the instructional goals to conform to a particular standard, then the system designed for extensibility should have a basic module for evaluation which could be extended to add/remove features required as per new accreditation requirements. 

In essence, variability is a broad concept and can range from user needs, market segments, customer profiles which can be addressed through a wide variety of artifacts that are generated throughout the software development life cycle.
In this paper, we attempt to address variability in instructional design as well as in software through patterns and software product lines. 

The core idea of POSA is to create software architecture using patterns such that these patterns can address variability and map to patterns in POID. Architecture patterns can further consist of design patterns with each of them addressing variability at different levels of granularity. In this section, we briefly provide examples for both architecture and design patterns. 

%\subsubsection{Architectural Patterns}

We illustrate the idea of POSA through a commonly used pattern called Model-View-Controller (MVC). Kranser and Pope have described MVC for interactive user interface applications \citep{krasner1988description}. The key purpose of MVC pattern is to facilitate separation of the interactive application into three parts or components to efficiently address changing requirements. Models primarily represent the underlying application domain knowledge and act as core structure for Views and Controllers. Instructional design is the underlying domain in this paper and as such forms the basis for models. The representation of this model itself can vary based on how the domain is modeled. As emphasized by Kranser and Pope, the focus should be on modeling specific information about the application domain such that it can drive the other two parts. Views primarily focus on the user interface, graphical elements and what the users view as part of the system. A typical user interface consists of hierarchical views and the data for these views is fed from the model behind them. This is quite useful as a model can have many views associated with it facilitating variability from a user interface perspective. Consider a scenario where an \textit{e}Learning System uses the same model but can be viewed using several user interface metaphors. The variabilities can range from simple color or themes to complex modifications emerging from instructional design. The order of these views itself can be a source of variability. Controller is the third component of MVC pattern that acts as the interface between models and views. 
%Two primary uses of MVC pattern are reusability and flexibility

%The model contains the underlying classes whose instances are to be viewed and manipulated 
%The view contains objects used to render the appearance of the data from the model in the user interface 
%The controller contains the objects that control and handle the user’s interaction with the view and the model
%The Observable design pattern is normally used to separate the model from the view 

It is not uncommon to integrate several architectural patterns while designing a software application to address varied requirements \citep{buschmann2007pattern}. For example, Figure §\ref{fig:ch3LayeredMVCPattern} shows a simple overview of how MVC pattern can be integrated into a layered architectural pattern. Here the user interface, business logic and data are separated into three layers Presentation, Business and Domain. MVC pattern is spread across \textit{presentation} and \textit{business} layers. The MVC pattern itself is a composed pattern consisting of several design patterns and can be implemented in several ways leading to variations like Hierarchical MVC, PVC \citep{karagkasidis2008developing}. Several design patterns are used to implement MVC and its variations. The model part of the MVC pattern is part of the domain strongly mapped to data parts of the domain i.e., pattern-oriented instructional design. A simple way to vary the application is to change these data parts in the POID without changing the interfaces exposed to POSA. Even MVC itself uses \textit{Observer} pattern to notify views and controllers, providing a variation point. Composite pattern is commonly associated for constructing user interface elements in a hierarchical way and exposing only the top level view for the entire architecture leaving flexibility to change user interface elements. \textit{Strategy} pattern is commonly used to alter between different controllers and use concrete strategies at different points of time to facilitate different behaviors for different triggers in the software application. One key difference in this layered MVC architecture as shown in Figure §\ref{fig:ch3LayeredMVCPattern} is the use of components instead of classes emphasizing the need to model every class as a component with explicit interfaces to facilitate variability. The domain layer emphasizes the strong role of domain in this architecture than just traditional databases. The data is mapped to patterns in the domain as this can allow traceability of changes from domain to software. A 10-step process for MVC pattern \citep{buschmann1996pattern}  along with potential variabilities is shown below: 

\begin{figure*}
	\centering
	\includegraphics[width=1\linewidth]{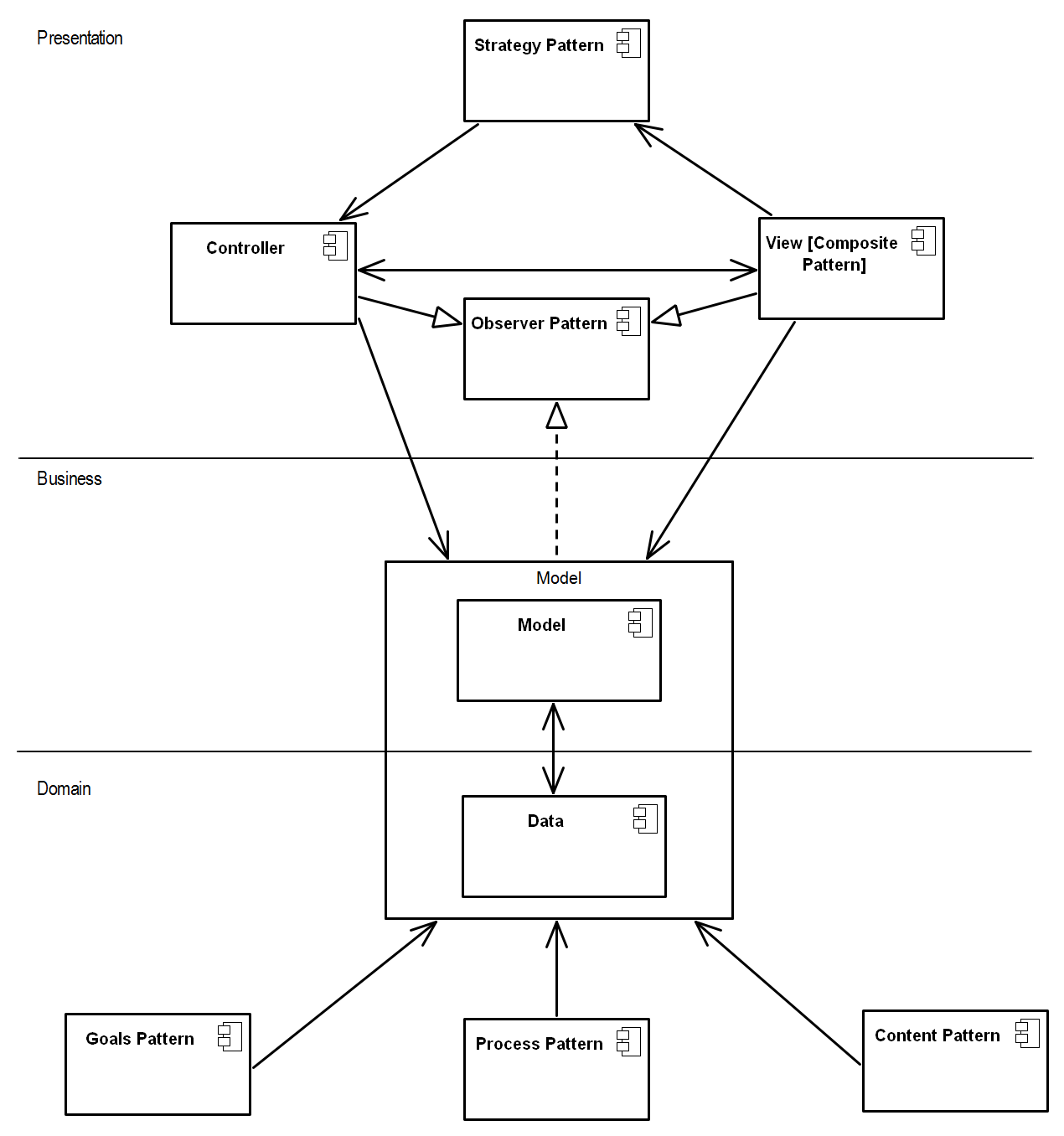}
	\caption{Layered Model View Controller Pattern}
	\label{fig:ch3LayeredMVCPattern}
\end{figure*}

\begin{enumerate}
	\item Separate human-computer interaction from core functionality \newline
	\textit{inputs, output behaviours, accessor functions}
	\item Implement the change-propagation mechanism \newline
	\textit{publish-subscribe pattern, specific implementations}
	\item Design and implement the views \newline
	\textit{appearance of views, display procedures, parameterized views, multiple draw methods}
	\item Design and implement the controllers \newline
	\textit{specific behaviours for user actions, event handling}
	\item Design and implement the view-controller relationship \newline
	\textit{initializations for which factory method pattern could be used, hierarchy of views and controllers}
	\item Implement the set-up of MVC \newline
	\textit{initializations, events}
	\item Dynamic view creation \newline
	\textit{components for managing views}
	\item Pluggable controllers \newline
	\textit{different controllers}
	\item Infrastructure for hierarchical views and controllers \newline
	\textit{composite pattern, chain of responsibility pattern}
	\item Further decoupling from system dependencies \newline
	\textit{bridge pattern, higher levels of abstraction}
\end{enumerate}

Today, there are several web frameworks such as \textit{Django}, \textit{Symfony}, \textit{Rails} that are based on MVC pattern but most importantly implement their own variation, addressing specific requirements. Most of the variations are possible by tweaking one or more of the steps in the above process. For example, there could be hundred controllers in a large scale application providing variability or the need to create varied interaction themes with users can be addressed by tweaking step 9. According to \citep{buschmann2007pattern},  MVC pattern references twelve design patterns such as \textit{Facade}, \textit{Observer} and so on. 

POSA can also be designed by integrating several design patterns. Gamma et al. list 23 design patterns and their relationships in a single diagram \citep{gamma1994design} and Zimmer provides a succinct classification of relationships between various design patterns into three layers: [design patterns specific to an application domain], [design patterns for typical software problems] and [basic design patterns and techniques] \citep{zimmer1995relationships}. To summarize, the core essence of this paper is to show how varying different aspects in patterns can facilitate variability needed for scale and variety in educational technologies. 

In the next section, we discuss an implementation of our approach that uses patterns as the base for instructional designs and educational technologies. 

\section{Implementation}
\label{sec:ch3Implementation}

Owing to the nature of our research work closely connected to the idea of solving societal challenges using computing, the implementation of our research work is made available at \url{http://rice.iiit.ac.in} and the work has been transferred to National Literacy Mission Authority , Government of India for further proliferation. The  mobile version of the generated software is deployed on Google Play Store and is available at \url{https://play.google.com/store/apps/details?id=iiit.rice.al.telugu}. In addition, our work is also listed in the official websites of Department of Adult Education of Government of Telangana at \url{http://tslma.nic.in/} and State Resource Center, Government of Telangana at \url{http://srctelangana.com/}. In this section, we will provide an overview of our implementation. 

%The goal of this paper was to create educational technologies based on instructional design and to support a variety of instructional designs. The patterns-based approach presented in this paper, especially  

Our initial implementation was a manual approach based on patterns in instructional design. We have designed an authoring tool called \textit{EasyAuthor} that helps non-technical teachers to create educational content \citep{Chimalakonda2013b} based on pedagogy patterns through our IDont framework \citep{Chimalakonda2013d}. This tool has wizards that help teachers to create different aspects of ID (context, goals, process, content, evaluation and environment) and eases the authoring process. Each of these wizards are based on patterns for corresponding aspects of the instructional design. This essentially connects different components of educational technologies to instructional design. 

However, a critical need is to create a variety of authoring tools based on a variety of instructional designs. For example, there is a need for \textit{EasyAuthor1} mapping to an  \textit{InstructionalDesignModel1}, \textit{EasyAuthor2} for \textit{InstructionalDesignModel2} and so on. To address this need of automating the development of \textit{EasyAuthor(N)} tools for varied \textit{InstructionalDesignModel(N)}, we have developed a software product line approach based on ontologies \citep{chimalakonda2017software}. 

\begin{figure*}
	\centering
	\includegraphics[width=1\linewidth]{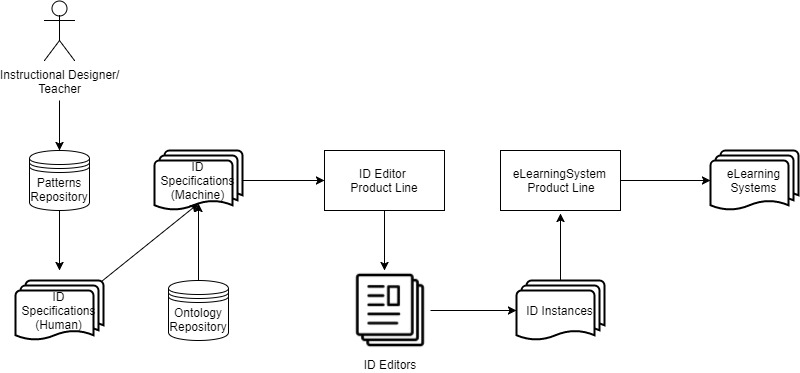}
	\caption{Overview of Implementation of Pattern-Oriented Approach}
	\label{fig:ch3Implementation}
\end{figure*}

Figure §\ref{fig:ch3Implementation} shows an overview of the implementation of our approach. Here, an instructional designer first decides the instructional design model for a set of courses that have to be taught. For example, the instructional designer may choose to use ``project based learning" as a strategy to teach several courses. He creates a set of patterns for aspects such as \textit{goals}, \textit{process}, \textit{content} either from scratch or by customizing existing patterns to suit the specific instructional design model. However, this model is mainly aimed at different stakeholders such as teachers, instructional designers, evaluators, policy makers and so on. These instructional design specifications have to be converted to specifications that are machine-processable such that automation is possible. To achieve this, we have used ontologies to concretely represent the patterns \citep{chimalakonda2017software}. These specifications are read by a tool called \textit{ID Editor Product Line} that semi-automatically generates an ``\textit{ID Editor}" that essentially allows a teacher to create a specific instance of instructional design. This \textit{ID Editor} is driven by the idea of connecting patterns in Pattern-Oriented Instructional Design and Pattern-Oriented Software Architecture. The \textit{ID Instances} are then used as input to another product line for generating \textit{e}Learning Systems. The details of how these product lines are implemented is beyond the scope of this paper and are detailed in the thesis \citep{chimalakonda2017software}. Further details on implementation and source code of implementation are available through \url{http://rice.iiit.ac.in}.

%Here, we show a few screenshots of 

%GURU is another experimental tool based on MIT Scratch that uses block structured drag-n-drop interface design to help non-technical teachers in creating ID [20]. While the interface was quite intuitive to teachers and is based on patterns, the very nature of MIT Scratch (no exposed APIs) made automation unfeasible. Whereas, EasyAuthor tool exploits the notion of patterns in both POID and POSA allowing us to automatically generate the tools themselves. We have also devised a software product lines approach that automates the generation of EasyAuthor(N), which is not the main focus of this paper. In short, this automation was possible only because of the systematic patterns-based approach employed in both POID and POSA.

\section{Conclusions \& Future Work}
\label{sec:ch3Conclusion}

%We emphasized that lack of instructional design as a basis for educational technologies and supporting a variety of instructional designs are two major research challenges for facilitating reuse during design of educational technologies. 

We emphasized that lack of instructional design base is a critical challenge for design of educational technologies. So is the challenge of facilitating reuse and
supporting a variety of instructional designs. To address these concerns, we presented a patterns-based approach that integrates patterns in instructional design and educational technologies. This approach has its roots in fundamental architecture principles in software engineering. Based on these principles, we presented an architecture that integrates \textit{Pattern-Oriented Instructional Design} that is driven by instructional design methodologies and Pattern-Oriented Software Architecture that drives the design of educational technologies. The essence of our approach is to systematically model different aspects of instructional design (\textit{goals}, \textit{process}, \textit{content}) using \textit{patterns} such as \textit{GoalsPattern},  \textit{ProcessPattern} (plays, acts, scenes, instructions) and \textit{ContentPattern} (facts, cases, rules, models and theories). We demonstrated the application of our approach to model patterns in adult literacy case study in India. We then provided an implementation of our approach that generates instructional design authoring tools based on patterns. Finally, we see this paper as a major research direction that addresses challenges in design of educational technologies through solutions in software engineering. 

This paper is an attempt to integrate research on patterns and patterns-based approaches in software engineering and instructional design to facilitate design of educational technologies that have a pedagogical basis and support systematic reuse. There many ways of improving our work. Firstly, there is no way to claim that the patterns we presented are the only possible patterns in instructional design nor they are comprehensive. In each category of patterns such as goals, process and so on, there is possibility of a variety of patterns in multiple contexts. In addition, there are still many manual steps in the pattern life cycle. While using patterns facilitates reuse, we see the following future research directions:

\begin{itemize}
	\item Discovering new patterns is a critical future direction to support the breadth of instructional designs, so are the mechanisms to support evolution of existing patterns. 
	\item Modeling notations for representing patterns and supporting techniques and tools for handling patterns throughout life cycle.
	\item Formal approaches for composing patterns either within the domain or software or between domain and software, validating assembly of patterns and their integration. 
	\item Applying our approach to beyond adult literacy for schooling and other forms of education.
\end{itemize}

% if have a single appendix:
%\appendix[Proof of the Zonklar Equations]
% or
%\appendix  % for no appendix heading
% do not use \section anymore after \appendix, only \section*
% is possibly needed

% use appendices with more than one appendix
% then use \section to start each appendix
% you must declare a \section before using any
% \subsection or using \label (\appendices by itself
% starts a section numbered zero.)
%

\section*{Acknowledgements}
We would like to thank TCS for providing us with initial inputs for this work, NLM for taking our work forward to create national impact, Government of Telangana for being one of the first adoptors of our technologies and all funding agencies for supporting several international research travels.

%\begin{IEEEbiographynophoto}{Jane Doe}
%Biography text here.
%\end{IEEEbiographynophoto}

% You can push biographies down or up by placing
% a \vfill before or after them. The appropriate
% use of \vfill depends on what kind of text is
% on the last page and whether or not the columns
% are being equalized.

%\vfill

% Can be used to pull up biographies so that the bottom of the last one
% is flush with the other column.
%\enlargethispage{-5in}
%\bibliographystyle{IEEEtran}
%\bibliographystyle{apalike}
\bibliographystyle{apacite1}
\bibliography{MyThesis1}

\end{document}